# Non-Monotonic Concentration Dependence of the Electro-Phoretic Mobility of Charged Spheres in Realistic Salt Free Suspensions

Denis Botin[a], Felix Carrique[b], Emilio Ruiz-Reina[c] and Thomas Palberg[§,a]

[a] *Institute of Physics, Johannes Gutenberg University, 55128 Mainz, Germany*

[b] *Institute Carlos I for Theoretical and Computational Physics (iC1), Departamento de Física Aplicada I, Facultad de Ciencias, Universidad de Málaga, Campus de Teatinos, 29071 Málaga, Spain*

[c] *Institute Carlos I for Theoretical and Computational Physics (iC1), Departamento de Física Aplicada II, Escuela de Ingenierías Industriales, Universidad de Málaga, Campus de Teatinos, 29071 Málaga, Spain*

[§] *correspondence should be sent to palberg@uni-mainz.de*

**Using super-heterodyne Doppler velocimetry with multiple scattering correction, we extend the optically accessible range of concentrations in experiments on colloidal electro-kinetics. We here measured the electro-phoretic mobility and the DC conductivity of aqueous charged sphere suspensions covering about three orders of magnitude in particle concentrations and transmissions as low as 40%. The extended concentration range for the first time allows the demonstration of a non-monotonic concentration dependence of the mobility for a single particle species. Our observations reconcile previous experimental observations made on other species over restricted concentration ranges. We compare our results to state of the art theoretical calculations using a constant particle charge and the carefully determined experimental boundary conditions as input. In particular, we consider so-called realistic salt free conditions, i.e. we respect the release of counter-ions by the particles, the solvent hydrolysis and the formation of carbonic acid from dissolved neutral $CO_2$. We also compare to previous results obtained under similarly well-defined conditions. This allows identification of three distinct regions of differing density dependence. An ascent during the built up of double layer overlap which is not expected by theory, an extended plateau region in quantitative agreement with theoretical expectation based on a constant effective charge and a sudden decrease which occurs way before the expected gradual decrease. Our observations suggest a relation of the non-monotonic behavior to a decrease of particle charge, and we tentatively discuss possibly underlying mechanisms.**

## Introduction

Electro-kinetic effects are of a great importance for many physicochemical, biological and industrial issues, ranging from signalling in nervous fibers over suspension processing and stability to desalination of a sea-water. The concentrations of colloidal particles in such studies vary over several orders of magnitude, depending on the focus of interest. The density dependence of electro-kinetic effects is therefore of great fundamental and practical interest.

The theory of electro-kinetic effects, the so-called Standard Electro-Kinetic Model is well established, and IUPAC recommendations give standard protocols for measurement procedures and conversion of measured mobilities to ζ-potentials or charges [1]. Few effects escaping mean field treatment (e.g. charge reversal observed in electrolytes with multivalent ions [2, 3, 4]) can be approached by primitive level calculations and simulations including full hydrodynamics [5]. In general, the Standard Electro-Kinetic Model comprises a set of coupled equations that can be solved under given boundary conditions either analytically or numerically [6, 7, 8, 9]. Both requires the previous resolution of the Poisson–Boltzmann equation (PB) within some cell-model calculation [10, 11, 12] to provide the particle charge, which is then kept constant in the electro-kinetic calculations. If necessary, further account can be made for charge regulation [13] and/or the double layer storage capacity for electrolyte ions [14]. Recent developments of the Standard Electro-Kinetic Model also include realistic salt free conditions, i.e. they account for counter-ions released by the particles themselves, for solvent hydrolysis and the formation of carbonic acid from dissolved neutral $CO_2$ [15, 16].

An important unsolved problem, however, is the particle concentration dependence of the electro-phoretic mobility. Concentration is expressed in terms of either the volume fraction $\Phi$ or the particle number density $n$. Both are related as $\Phi = n(4\pi/3)a^3$ with $a$ being the particle radius. In the context of electro-kinetics, the number density is the more appropriate parameter, as it directly relates to the electrolyte concentration *via* the number of released counter-ions per charged particle. For a constant charge, the Standard Electro-Kinetic Model expects an extended plateau at low concentrations and a gradual, approximately logarithmic, decrease at large concentrations. However, already very early, Dunstan, Rosen and Saville reported some indications for a maximum of the mobility at low volume fractions and low salt concentrations, but at the same time, they stressed the strong dependence of their observations on sample conditioning [17]. This necessitated the implementation of advanced conditioning methods [18] providing a flexible adjustment of particle and electrolyte concentration under precise in situ conductometric control. Subsequent studies on small, strongly interacting particles reported a mobility plateau followed by a decrease towards large concentrations [19, 20], i.e. in line with theoretical expectation [15] and simulations on the primitive level [21]. This type of decrease starts at the onset of counter-ion domination and is therefore traced back to the counter-ion contribution to the overall salinity. In addition, also a much steeper decrease was recently observed, albeit again under not well-controlled conditioning [22].

Interestingly, in early work, the decrease was also suggested to possibly relate to the onset of crystallization (the freezing transition [23]) and thus correspond to an increase of increased suspension viscosity [24]. For highly charged spheres in low salt aqueous suspension, the freezing transition may be found at very low volume fractions in the permille range [25]. Across the freezing transition and in the crystalline phase, multiphase flow, shear banding and other non-linear flow types may occur upon application of shear [26]. These also are frequent in electro-kinetic experiments that are conducted in closed cell geometries [27], where shear results quite naturally from the electro-osmotic solvent flow along the cell walls. In this case, special measures have to be taken to separate solvent flow and particle motion against the solvent. We could show that taking velocity averages across the complete cell cross-section averages out any underlying solvent motion and allows extracting electro-kinetic mobilities in a reliable way even in the presence of flow instabilities [19, 20, 28]. The obtained electro-phoretic mobilities varied smoothly over the transition, and the location of freezing did in general not coincide with the onset of the decrease. Therefore, the decrease of mobility is not related to an increase of suspension viscosity or to crystallinity. This, in fact, is also supported by theory, as the particle mobility relates to the slip velocity between solvent and particle surface rather than to the relative motion of particles in the bulk of the suspension. For an extended discussion of colloidal crystal electro-phoresis, the interested reader is referred to [27]

By contrast, other studies on more dilute low-salt suspensions of large particles showed a slow *increase* of mobility with increased concentration followed by a plateau [24, 29, 30]. This type of somewhat counter-intuitive number density dependence was also found at elevated salt concentrations for silica particles [31] and for surfactant stabilized latex particles [28]. Such an increase is not at all expected from the standard electro kinetic model, which at low concentrations predicts a constant mobility connecting monotonically to the decrease at larger concentrations. The concentration dependent experimental investigations are complemented by studies on isolated particles in either aqueous [32, 33, 34] or organic solvent [35]. There, the observed mobilities were found to be independent of particle radius and of very small magnitude. The first finding was attributed to charge renormalization which compensates for neglecting finite ion sizes in mean field [10, 11, 12, 36, 37]. The second finding implies very low effective charges, much smaller than typically observed for the same particles in conductivity experiments at elevated concentration [38] and also significantly lower than expected from theory [39]. Low limiting mobility, increase and plateau have been observed in systems of different surface chemistry, different sizes and various background electrolyte concentrations $10^{-6} \leq c_S \leq 10^{-2}$ mol $L^{-1}$ [30, 31, 28, 32, 33, 35]. However, no clear correlation of experimental boundary conditions to the cross-over from increase to a constant value was so far observed, although several authors suggested a possible correlation to the onset of double layer overlap and/or of structure formation [29, 24, 40].

Clearly, measurements over an extended range of concentrations are highly desired to capture the density dependence as completely as possible. All earlier studies, however, were restricted in their range of accessible particle concentrations by the loss of signal at low concentrations and by multiple scattering at large concentrations. In the present paper we present the first systematic study over a sufficiently large range of volume fractions to catch the density dependence centered about a broad plateau maximum. This includes highly dilute samples, samples of fluid order, and crystalline samples. To avoid signal loss at low concentrations we employ comparably large spheres. To cope with multiple scattering, we use a recently introduced variant of Super-Heterodyne Laser Doppler Velocimetry (SH-LDV) [22]. In integral small angle configuration, this instrument is capable to yield multiple scattering (MS) free electro-phoretic mobilities over several orders of magnitude in concentration. Moreover, integral measurements allow for simultaneous measurements of electro-osmotic mobilities [41].

Deviations between theoretical expectation and experimental observation may easily result from ill-controlled experimental boundary conditions or relate to some sample specific surface chemistry. A a careful particle and boundary condition characterization from independent experiments is therefore indispensable. In our earlier study using SH-LDV, we worked with initially deionized samples that were subsequently exposed to ambient air. The preliminary results supported the suggestion of a mobility maximum but no precise control of the electrolyte concentration was applied at that stage. In the present study, we therefore combine SH-LDV with improved conditioning control, and we complement our mobility experiments by simultaneous measurements of the DC conductivity and the suspension turbidity.

In what follows we first describe the experimental conditioning protocols and the system characterization in some detail. This will generate reliable input for the theoretical calculations. We then give an outline of SH-LDV with some emphasis on performance under conditions of multiple scattering. We illustrate the applied multiple scattering correction procedures and show our data evaluation. The theoretical section introduces the Standard Electro-Kinetic Model used to calculate predictions for the concentration dependence of the electro-phoretic mobility of the experimental system. We then present out experimental results and compare them to theoretical calculations based on the results of system characterization. In the discussion, we first compare the experimental results to selected previous findings obtained under similarly well-defined boundary conditions. We demonstrate that the non-monotonic density dependence of the electro-phoretic mobility reconciles these apparently conflicting observations made over restricted concentration ranges. Next we discuss both electro-phoretic and electro-osmotic mobilities in view of the theoretical predictions made for realistic salt free conditions. In the plateau region, we can give a consistent interpretation of our experimental mobilities that reveals a constant saturated $CO_2$ background and a constant effective charge. Outside the plateau region we find a significant drop of the $\zeta$-potential and hence the effective charge both towards smaller and towards larger $n$. We also compare our results to selected previous findings, which allows suggesting possible mechanisms for charge reduction. We close with a short outlook to ongoing experiments.

## Materials and methods

### Sample conditioning and characterization

We used commercial polystyrene latex spheres, stabilized by carboxyl surface groups (lab code PS260(mP), Lot PS-F-L2208, microParticles GmbH, Berlin, Germany). We obtained hydrodynamic particle diameters of $2a_h$ = (268±2) nm by standard dynamic light scattering (DLS) and geometric diameters of $2a_{SLS}$ = (254±4) nm from the static form factor measurements using static light scattering (SLS). The polydispersity index given by the manufacturer is *PI* = 3.1%.

All measurements were conducted in a climatized lab at an ambient temperature $\vartheta$ = (25.2±0.2) °C Dispersions were conditioned in a closed circuit system [18] The sample is continuously deionized, while passing through a mixed bed ion-exchange resin filled column (IEX, Amberlite K306, Carl Roth GmbH  Co. KG, Karlsruhe, Germany). The IEX column can be bypassed after attaining fully deionized conditions. An integrated DC conductometric cell (electrodes LR325/01 bridge LF340i, $f_{AC}$ = 400 Hz, WTW, Germany) allows monitoring the residual ion concentration and detecting of impurity leaks. The leakage rates of the circuit were determined from-measurements with pure water.

Deionizing for sufficiently long times (e.g. 15 min for 30 ml of suspension) yields the measured conductivity of $\sigma_{min}$ = (60-65) nS cm$^{-1}$, corresponding to a concentration of carbonate ions of $1.5\times10^{-8}$ mol L$^{-1}$ or a residual concentration of dissolved $CO_2$ of $6.5\times10^{-8}$ mol L$^{-1}$. The typical duration of an electro-kinetic experiment is 15 min, during which leakage of the $CO_2$ from the ambient air into the circuit occurs. We further checked that the conductivity increase in a measuring cell isolated from the rest of the circuit by two-way valves is about one order of magnitude slower than in the complete circuit with its many tubing connections. In isolated cells, $CO_2$ leakage from external sources is therefore negligible as compared to the increase of the minimum conductivity by adding particles. However, at finite number density, *n*, we observe the conductivity rise to a significantly larger extend than in pure water. This shows that an additional source of $CO_2$ must be present. We attribute it to $CO_2$ stored by adsorption onto the particle surface [42]. To ensure well-deionized conditions during the experiments, we therefore isolate the cells after circulating the suspensions a few minutes at constant minimal conductivity. The obtained minimal conductivities are displayed in Fig. 1a as a function of *n* as calibrated from the static light scattering on crystalline samples.

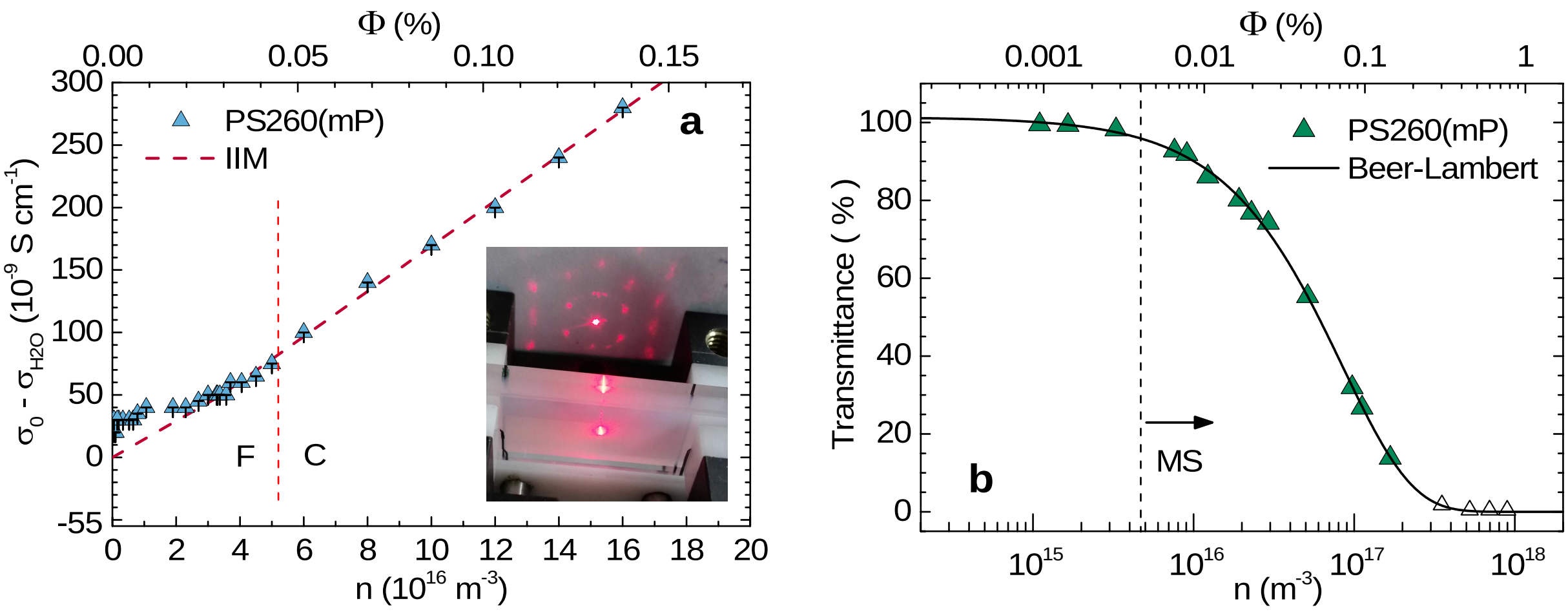

Fig. 1: Suspension characterization. Lower scales: number densities, upper scales: volume fractions. a) Concentration dependence of the minimal low frequency conductivity of realistic salt free PS260(mP). Conductivity data are corrected for the nominal contribution of water hydrolysis $\sigma_{H2O}$ = 55 nS cm$^{-1}$ at $\vartheta$ = (25.2±0.2)°C. The freezing concentration $n_F$ is indicated as vertical dashed line. Starting from $n \approx 3\times10^{16}$ m$^{-3}$, data show a linear $n$-dependence. The dashed line is a fit of the independent ion migration model [43, 44] returning $Z_\sigma = -2500 \pm 30$. Inset: Laue pattern of an oriented bcc crystal forming in the electro-kinetic cell for $n \geq n_F = (4.6\pm0.2) \times 10^{16}$ m$^{-3}$ ($\Phi \geq \Phi_F = 4\ 10^{-4}$). Image taken at $n = 8.5 \times 10^{16}$ m$^{-3}$. Note further the pronounced halo inside the electro-kinetic cell resulting from multiple scattering. b) Relative transmitted intensity of PS260(mP) at $\lambda = 633$ nm normalized to the cell thickness, $2d$. Solid line is a fit of the Lambert-Beer attenuation law. The vertical dashed line denotes the onset of multiple scattering at $n \approx 5\times10^{15}$ m$^{-3}$, i.e. $\Phi \approx 5\times10^{-5}$ as visible in standard dynamic light scattering. In SH-LDV, we observe a substantial influence of multiple scattering on the spectra for $n \geq 2\times10^{16}$ m$^{-3}$.

Data show a linear increase at large $n$, but as expected, the conductivity levels off towards low $n$. From static light scattering, we observe the transition to the crystalline state for our highly charged spheres at $n_F = (4.6\pm0.2) \times10^{16}$ m$^{-3}$, i.e. at a volume fraction of $\Phi_F = 4\ 10^{-4}$ (vertical dashed line). Note, that the conductivity is not influenced by this freezing transition. Instead, for $n > 3\times10^{16}$ m$^{-3}$, data are well described by the linear dependence expected within Hessinger's model of independent ion migration [43, 44] which returns an effective charge of $Z_{\mathrm{eff}} = -2500\pm30$. Hessinger's model determines the effective number of freely moving counter-ions which typically is found to be in a good agreement with the values of the renormalized charge derived from cell model calculations [10, 12, 36, 39]. For a comparison of these and other charge numbers using Alexander's renormalized charge as a reference, see [37]. Assuming that the bare charge number is large enough to reach the renormalization limit [36], the amount of renormalization can be estimated from the dimensionless scaling parameter $A$ via $Z_{\mathrm{eff}} = Aa(1+\kappa a)/\lambda_B$ [35]. Estimating $A$ with the Bjerrum length in water of $\lambda_B = 0.7$ nm and a Debye-Hückel screening parameter of $\kappa \approx 1.5\times10^{6}$ m$^{-1}$, we here obtain $A = 7.9$, in good agreement with previous estimates based on the effective electro-kinetic charges of other highly charged latex spheres [37, 38]. Interestingly, the scaling relation also holds for effective charges obtained from electro-kinetic experiments at low particle concentration or on isolated particles [33, 34, 35], albeit with much smaller than theoretically expected values [39].

For turbidity measurements, a laser beam ($\lambda = 633$ nm) passes through a rectangular optical cell, and the light transmittance is measured by a power meter (PM, LabMax-TO, Coherent Inc., USA). Turbidity experiments were conducted at the background electrolyte concentration adjusted to $c = 6.3\times10^{-6}$ mol L$^{-1}$ to obtain fluid ordered samples. Results are displayed in Fig. 1b. The transmitted light obeys the Beer-Lambert attenuation law $-\ln(I/I_0) = 2d\ n\ \sigma_{633}$ [45]. Here, $I$ and $I_0$ are the transmitted and the incident intensity, $d$ is the cell thickness, A fit using a refractive index of $\nu = 1.59$ returns an attenuation cross section $\sigma_{633} = (6.5\pm0.4)\times10^{-3}$ µm$^2$. This value corresponds to a particle diameter of $2a = (247\pm16)$ nm in good agreement with the optical radius from SLS.

**Electro-kinetic measurements**

Electro-kinetic experiments were conducted in the range of linear attenuation, corresponding to moderate MS, i.e. less than 5 scattering events per photon [22]. We used U-shaped quartz cells with rectangular optical cross-sections (EL10 by Rank Bros., Bottisham, Cambridge, UK or replica by Lightpath Optical Ltd., Milton Keynes, UK) and various thicknesses 2*d*. The cells were cleaned by heating in a flame of a gas burner for some 15 min, letting to cool and successive sonication for 60 min at 35°C in 2% alkaline detergent solution (Hellmanex III, Hellma Analytics). They were further rinsed with milli-Q water and dried in a nitrogen stream. We used platinized platinum electrodes. For each cell, the effective electrode distance l was calibrated using standard KCl solutions and is on the order of some 7 cm. To avoid accumulation of particles at electrodes, *AC* square-wave fields of $E = 10 – 45$ V cm$^{-1}$ and $f_{AC} = (0.02 – 0.1)$ Hz were utilized.

Super-heterodyne Laser Doppler Velocimetry (SH-LDV) is a recently introduced variant of heterodyne Laser Doppler velocimetry. The detailed description and characterization of the experimental setup, together with the corresponding scattering theory has been given elsewhere [22, 40, 41]. It is based on a standard heterodyne light scattering or Doppler velocimetry [46], but uses a super-heterodyning approach of side band modulated spectroscopy [47] or Phase Analysis Light Scattering (PALS) [48, 49] allowing isolation of the desired heterodyne signal component from the homodyne part and low frequency noise. Operating in the integral mode [50], it averages over the complete cell cross section at mid-cell height. The super-heterodyne (*shet*) spectrum is then directly proportional to the particle velocity distribution in that plane. Thus, the velocity distribution is measured in a single run, instead of tedious point-by-point measurements, and both the electro-phoretic and the electro-osmotic velocities can be obtained simultaneously [41]. Moreover, any deviations from the ideal parabolic flow profile are seen immediately. As a small angle scattering experiment, it delivers spectra of suspensions at rest or under electro-kinetic flow irrespective of particle concentration and suspension structure [40]. The most important advantage, however, is the possibility to identify and correct for multiple scattering contributions to the Doppler spectra [22].

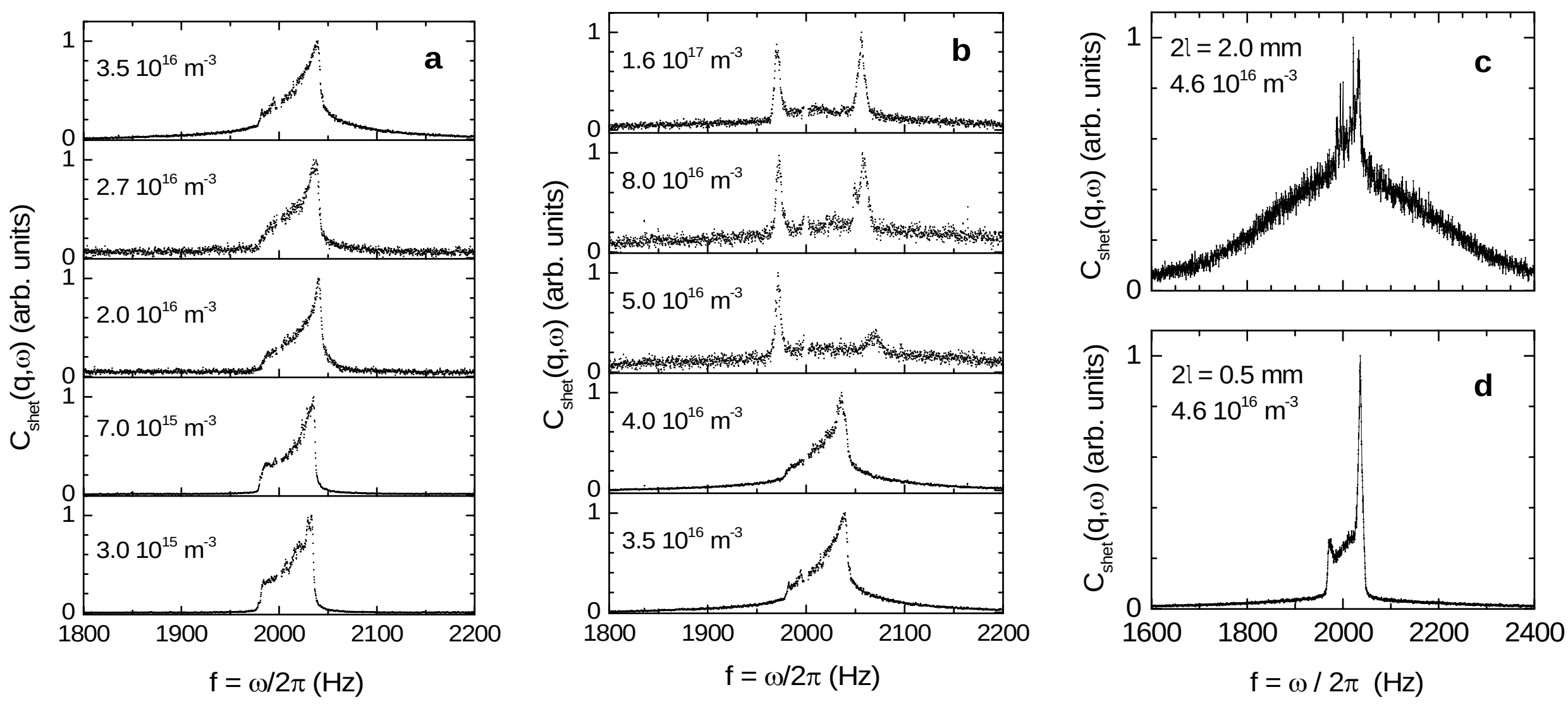

Fig. 2: Selected spectra normalized to maximum intensity. Applied voltage $U = 151$ V. a) and b): Evolution of spectral shape with increasing *n* as indicated. The cell depth (field strength) is $2d = 2$ mm ($E = 13.1$ V cm$^{-1}$)

in a) and the two lower spectra in b). The cell depth (field strength) is $2d$ = 0.5 mm ($E$ = 25.0 V cm$^{-1}$) for the upper three spectra in b). c) Disappearance of the signal in the MS background and spectral distortion due to flow instability at $n = 4.6 \times 10^{16}$ m$^{-3}$. d) The same suspension studied in a thinner cell shows stable stationary multiphase flow and reduced multiple scattering contribution. For details, see text.

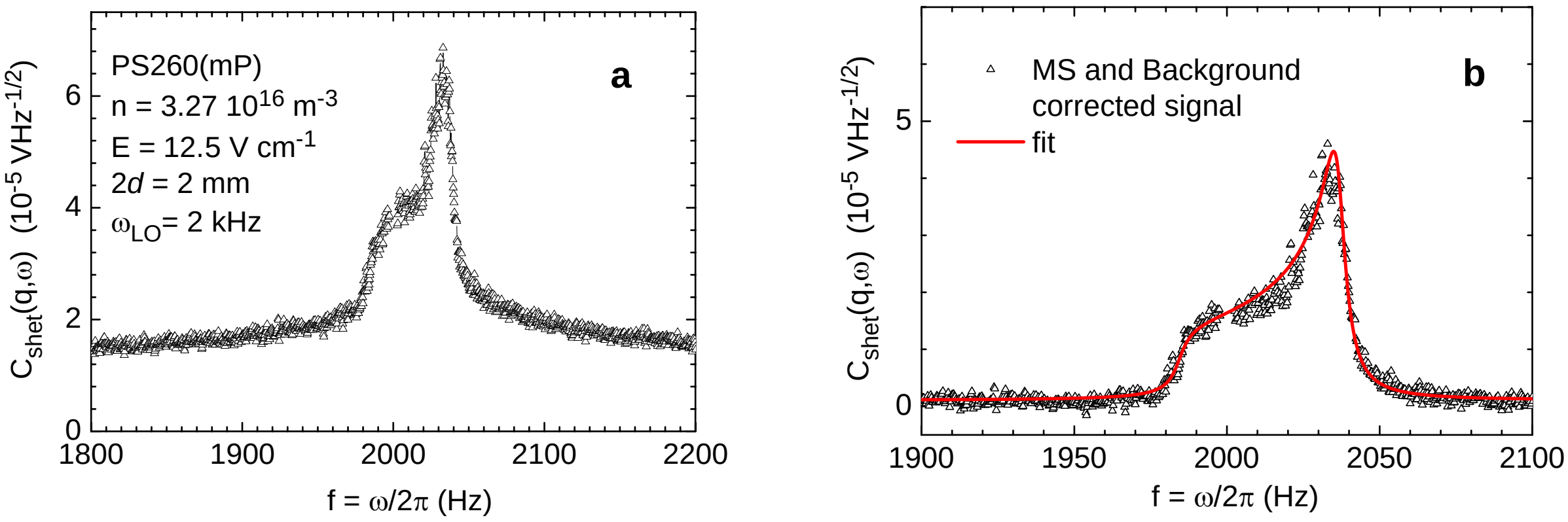

Fig. 3: Correction and evaluation of spectra. a) Example of a raw time averaged spectrum taken at $n = 3.27 \times 10^{16}$ m$^{-3}$, $E$ = 12.5 V cm$^{-1}$ and $2d$ = 2 mm. b) The same after subtraction of the MS Lorentzian and the static background. The solid line is a fit to the corrected data returning $v_{eo}$ = 91.25 µm s$^{-1}$; $v_{ep}$= 45.0 µm s$^{-1}$, and $D_{eff}$ = $5.96 \times 10^{-12}$ m$^2$s$^{-1}$.

In Fig. 2a and b, we display power spectra for increasing $n$. A broad multiple scattering contribution appears for $n \geq 2\times10^{16}$ m$^{-3}$ and gains in strength. It swallows the signal for $n \geq 4.5\times10^{16}$ m$^{-3}$ (Fig. 2c). Accidentally, here $n \approx n_F$, and a shear induced flow instability occurs, which strongly distorts the spectral shape [51, 52]. Using a thinner cell of $2d$ = 0.5 mm the flow instability gave way to stationary flow of counter-propagating crystals at the walls and in the cell center, separated by a broad fluid region [53]. Moreover, the amount of MS is reduced considerably (Fig. 2d). With further increasing $n$ the fluid gap is reduced forming a thin shear band at the largest investigated $n$ (upper three spectra in Fig. 2b), while application of larger fields widens the gap again (not shown).

All spectra shown were obtained averaging over some hundred time intervals restricted to one field direction, starting after the full development of electro-osmotic flow profile and ending shortly before the field reversal. Fig. 3a displays a raw spectrum of a fluid sample. In Fig. 3b, we subtracted the static background and the MS contribution as obtained by fitting a Lorentzian to the wings of the MS-signal [22]. We then fitted the remaining signal by standard procedures [40]. As a rule of thumb, for fluid samples the center of mass frequency of the spectrum corresponds to the electro-phoretic mobility, while the width of the asymmetric broadening is related to the cell geometry and the electro-osmotic mobility. In crystalline samples, we followed [20, 28] and varied the vertical location of the scattering plane (y-direction) during each measurement series at constant field strength. Then the electrophoretic mobilities derive as double average of particle velocities, $v_P$, over the complete cross section of the cell: $v_{ep} = \langle\langle v_P(x,y)\rangle_x\rangle_y$ . Note that this inhibits evaluation of electro-osmotic velocities. For each particle concentration, up to seven different field strengths were applied and velocities plotted versus $E$. The mobilities $\mu_{ep} = v_{ep}/E$ and $\mu_{eo} = v_{eo}/E$ were then obtained from linear fits to the data.

**Standard Electro-Kinetic Model for realistic salt free conditions**

We performed calculations within the Standard Electro-Kinetic Model for realistic salt free concentrated suspensions [16]. We have used a mean-field approach with a cell model scenario to account for the electro-hydrodynamic interactions between charged spherical particles in an average sense. This theoretical approach encompasses studies of many electro-kinetic phenomena in concentrated colloids as the electrophoresis [54, 55], electric permittivity [56], electro-viscous effect [57, 58] or electroacoustic phenomena [59, 60]. The cell model concept has been comprehensively covered in the review by Zholkovskij et al. [61]. Moreover, it has been profusely checked in different electro-kinetic experiments with charged particles in general electrolytes solutions, pure and realistic salt-free systems and non-polar environments [11, 15, 62, 63]. A spherical cell is composed by a particle of radius $a$ at its centre surrounded by a fluid layer with an outer radius $b$. The particle volume fraction $\Phi$ of the suspension determines the outer radius $b$ of the cell, as it is obeyed that the particle volume fraction calculated in the ambit of the cell coincides with that of the whole suspension according to Happel boundary condition [64]: $\Phi = (a/b)^3$. We admit that the suspension properties can be obtained from appropriate averages of local properties in the cell where also appropriate boundary conditions at the outer cell surface are imposed to manage with electro-hydrodynamic interactions. We characterize the particle by a surface charge density $\sigma$, and the solution contains added counter-ions released by the particles by dissociation of surface groups responsible for the particle charging process. Apart from the added counter-ions species, that will be assumed in this work as $H^+$, if the particles are negatively charged, or $OH^-$ if positively charged, there are also ions from different sources as $H^+$ and $OH^-$ from water dissociation, and $H^+$, bicarbonate anions $HCO_3^-$ and neutral carbonic acid molecules, $H_2CO_3$, from the atmospheric $CO_2$ contamination. According to the solubility and partial pressure of $CO_2$ in standard air at 25ºC, the concentration of $CO_2$ in water is $1.18\times10^{-5}$ mol $L^{-1}$. The chemical reactions in the realistic salt-free aqueous solution are:

$$\begin{aligned} &H_2O \underset{K_{-1}}{\overset{K_1}{\rightleftarrows}} H^+ + OH^- \\ &H_2CO_3 \underset{K_{-2}}{\overset{K_2}{\rightleftarrows}} H^+ + HCO_3^- \\ &H_2CO_3 \underset{K_{-3}}{\overset{K_3}{\rightleftarrows}} CO_2 + H_2O \end{aligned} \qquad (1)$$

$K_i$ and $K_{-i}$ ($i$=1, 2, 3) are the corresponding forward ($s^{-1}$) and backward ($m^3\,s^{-1}$) kinetic constants.

We will review here the main theoretical aspects of the electro-kinetic cell model for concentrated suspensions in realistic salt-free conditions in a static electric field **E**, which is the DC limit of the general electro-kinetic model of the dynamic electrophoretic mobility and dielectric response in AC electric fields [65]. The fundamental equations connecting the electrical potential $\Psi(\mathbf{r})$, the number density of each species, $n_j(\mathbf{r})$, their drift velocity $\mathbf{v}_j(\mathbf{r})$, the fluid velocity $\mathbf{v}(\mathbf{r})$, and the pressure $P(\mathbf{r})$ are:

$$\nabla^2\Psi(\mathbf{r}) = -\frac{\rho_{\text{el}}(\mathbf{r})}{\varepsilon_{\text{rs}}\varepsilon_0} \tag{2}$$

$$\rho_{\text{el}}(\mathbf{r}) = \sum_{k=1}^{3} z_k\, e\, n_k(\mathbf{r}) \tag{3}$$

$$\eta_{\text{s}}\nabla^2\mathbf{v}(\mathbf{r}) - \nabla P(\mathbf{r}) - \rho_{\text{el}}(\mathbf{r})\nabla\Psi(\mathbf{r}) = 0 \tag{4}$$

$$\nabla\cdot\mathbf{v}(\mathbf{r}) = 0 \tag{5}$$

$$\mathbf{v}_j(\mathbf{r}) = \mathbf{v}(\mathbf{r}) - \frac{D_j}{k_{\text{B}}T}\nabla\mu_j(\mathbf{r}) \quad (j=1,\ldots,4) \tag{6}$$

$$\mu_j(\mathbf{r}) = \mu_j^{\infty} + z_j e\,\Psi(\mathbf{r}) + k_{\text{B}}T\ln n_j(\mathbf{r}) \quad (j=1,\ldots,4) \tag{7}$$

$$\nabla\cdot\left[n_j(\mathbf{r})\mathbf{v}_j(\mathbf{r})\right] = \sigma_j(\mathbf{r}) \quad (j=1,\ldots,4) \tag{8}$$

Here, **r** is the position vector in a reference system fixed at the particle center. Also, $j$=1 stands for $H^+$, $j$=2 for $OH^-$, $j$=3 for $HCO_3^-$ and $j$=4 for the neutral species $H_2CO_3$. Eq. (2) is Poisson's equation, where the electric charge density $\rho_e$ is given by Eq. (3). $\varepsilon_{\text{rs}}$, $\varepsilon_0$ and $e$ are the relative permittivity of the solution, the permittivity of a vacuum and the elementary electric charge, respectively, and $z_k$ is the valence of the $k$-th ion. Eqs. (4) and (5) are the Navier-Stokes equations for an incompressible fluid flow of viscosity $\eta_s$ and mass density $\rho_s$ at low Reynolds number in the presence of an electrical body force. As we are just concerned with the linear response of the system to an applied electric field of low field-strength, the inertial term in the latter Navier-Stokes equation has been neglected, thus fulfilling the conditions of a Stokes incompressible fluid flow in the presence of an electrical body force. Eq. (6) derives from the Nernst-Planck equation for the flow of each $j$-th species. Eq. (7) stands for the electro-chemical potential where $\mu_j^{\infty}$ is its standard value, $D_j$ its diffusion coefficient, $k_B$ the Boltzmann constant and $T$ the absolute temperature. Eq. (8) is the continuity equation that allows for the conservation of chemical species even if they are generated o annihilated by chemical reactions. The latter aspects are represented by the following generation-recombination functions $\sigma_j(\mathbf{r})$, ($j$=1,…,4), for ions and neutral molecules involved in chemical reactions:

$$\begin{aligned}
\sigma_1(\mathbf{r}) &= \sigma_{H^+}(\mathbf{r}) = \left[K_1 n_{H_2O}(\mathbf{r}) - K_{-1} n_{H^+}(\mathbf{r}) n_{OH^-}(\mathbf{r})\right] + \left[K_2 n_{H_2CO_3}(\mathbf{r}) - K_{-2} n_{HCO_3^-}(\mathbf{r}) n_{H^+}(\mathbf{r})\right] \\
\sigma_2(\mathbf{r}) &= \sigma_{OH^-}(\mathbf{r}) = \left[K_1 n_{H_2O}(\mathbf{r}) - K_{-1} n_{H^+}(\mathbf{r}) n_{OH^-}(\mathbf{r})\right] \\
\sigma_3(\mathbf{r}) &= \sigma_{HCO_3^-}(\mathbf{r}) = \left[K_2 n_{H_2CO_3}(\mathbf{r}) - K_{-2} n_{HCO_3^-}(\mathbf{r}) n_{H^+}(\mathbf{r})\right] \\
\sigma_4(\mathbf{r}) &= \sigma_{H_2CO_3}(\mathbf{r}) = -\left[K_2 n_{H_2CO_3}(\mathbf{r}) - K_{-2} n_{HCO_3^-}(\mathbf{r}) n_{H^+}(\mathbf{r})\right] - \left[K_3 n_{H_2CO_3}(\mathbf{r}) - K_{-3} n_{H_2O}(\mathbf{r}) n_{CO_2}(\mathbf{r})\right]
\end{aligned}, \tag{9}$$

according to the procedure developed by Baygents and Saville for weak electrolytes [66].

The appropriate boundary conditions are:

$$\Psi_{\text{p}}(\mathbf{r}) = \Psi(\mathbf{r}) \quad \text{at } |\mathbf{r}|=a \tag{10}$$

$$\varepsilon_0\varepsilon_{\text{rs}}\nabla\Psi(\mathbf{r})\cdot\hat{\mathbf{r}} - \varepsilon_0\varepsilon_{\text{rp}}\nabla\Psi_{\text{p}}(\mathbf{r})\cdot\hat{\mathbf{r}} = -\sigma \quad \text{at } |\mathbf{r}|=a \tag{11}$$

$$\mathbf{v} = 0 \quad \text{at } |\mathbf{r}|=a \tag{12}$$

$$\mathbf{v}_j\cdot\hat{\mathbf{r}} = 0 \quad (j=1,\ldots,4) \quad \text{at } |\mathbf{r}|=a \tag{13}$$

$$\left\langle (\rho_{\rm m}\mathbf{v}') \right\rangle = \frac{1}{V_{\rm cell}} \int_{V_{\rm cell}} (\rho_{\rm m}\mathbf{v}')(\mathbf{r}')\, dV = 0 \tag{14}$$

$$\boldsymbol{\omega} = \nabla \times \mathbf{v} = 0 \qquad \text{at } |\mathbf{r}|=b \tag{15}$$

$$n_j(\mathbf{r}) = n_j^0(\mathbf{r}) \quad (j=1,\ldots,4) \quad \text{at } |\mathbf{r}|=b \tag{16}$$

$$\Psi(\mathbf{r}) = \Psi^0(r) - \mathbf{E}\cdot\hat{\mathbf{r}} \quad \text{at } |\mathbf{r}|=b \tag{17}$$

At the particle surface |**r**|=*r*=*a*, Eqs. (10) and (11) stand for the continuity of the electric potential and normal component of the displacement vector. Here, $\varepsilon_{\rm rp}$ is the relative permittivity of the particle (the sub-index p stands for the solid particle). Eq. (12) is the well-known non-slip condition for the fluid. Eq. (13) shows that the particle is impenetrable for ions and neutral molecules ($\hat{\mathbf{r}}$ is the radial unit vector pointing outwards to the particle surface). Eq. (14) states that the volume average of the macroscopic momentum per unit volume is zero [67]. $\rho_{\rm m}$ denotes the local mass density, which equals $\rho_{\rm s}$ or $\rho_{\rm p}$ when the solution or the solid particle are called in the volume integral over the cell volume $V_{\rm cell}$, respectively. $\mathbf{v}'$ and $\mathbf{r}'$ are the local velocity and position vector with respect to a fixed laboratory reference system. At the outer surface of the cell |**r**|=*r*=*b*, Eq. (15) follows the Kuwabara null vorticity condition for the fluid velocity [68], and Eqs. (16) and (17) are Shilov-Zharkikh-Borkovskaya's boundary conditions [69] for the concentrations of ions and neutral molecules, and of the electrical potential at the outer surface of the cell, respectively. An additional equation to close the problem comes from the equation of motion for the electro-neutral unit cell in the static case:

$$\int_0^{\pi} \left[\sigma_{rr}\cos\theta - \sigma_{r\theta}\sin\theta\right]_{r=b} \; 2\pi b^2 \sin\theta \; d\theta = 0 \tag{18}$$

Here, $\sigma_{rr}$ and $\sigma_{r\theta}$ are the normal and tangential components of the hydrodynamic stress tensor in spherical coordinates $(r, \theta, \varphi)$, with the polar axis ($\theta = 0$) parallel to the electric field.

The quantities with a superscript “0” in Eqs. (16) and (17) refer to equilibrium conditions (no applied electric field). The equilibrium electric potential and the equilibrium ionic concentrations are connected by the Poisson-Boltzmann equation (PB):

$$\begin{gathered} \frac{1}{r^2}\frac{d}{dr}\left(r^2\frac{d\Psi^0}{dr}\right) = \frac{d^2\Psi^0}{dr^2} + \frac{2}{r}\frac{d\Psi^0}{dr} = -\frac{\rho_{\rm el}^0}{\varepsilon_{\rm rs}\varepsilon_0} \\ \rho_{\rm el}^0(r) = \sum_{k=1}^{3} z_k\, e\, n_k^0(r) = \sum_{k=1}^{3} z_k\, e\, b_k \exp\left(-\frac{z_k\, e\,\Psi^0(r)}{k_{\rm B}T}\right) \end{gathered} \tag{19}$$

For each unknown coefficient $b_k$ the local concentration of the *k*-th ionic species at the outer surface of the cell where the equilibrium electric potential is chosen to be zero: $\Psi^0(b) = 0$ . The equilibrium boundary conditions to solve the PB equation are:

$$\begin{gathered} \left.\frac{d\Psi^0(r)}{dr}\right|_{r=b} = 0 \\ \left.\frac{d\Psi^0(r)}{dr}\right|_{r=a} = -\frac{\sigma}{\varepsilon_{\rm rs}\varepsilon_0} \end{gathered} \tag{20}$$

To obtain the unknown coefficients $b_k$, an iterative procedure must be used in connection with the equilibrium chemical equations and the electro-neutrality condition of the cell. A detailed study about the resolution method can be found in ref. [70, 71],

In addition, the electrophoretic mobility $\mu_{ep}$ can be obtained from the boundary condition in Eq. (14) for the velocity, as the velocity $\mathbf{v}'$ in the cell is the electro-phoretic velocity $\mu_{ep}\mathbf{E}$ when it refers to the solid particle [72]. We solved the electro-kinetic equations with the mentioned boundary conditions numerically using ODE Solver routines implemented in MATLAB©. For brevity, we omit technical aspects concerning the numerical and analytical procedures. We refer the interested reader to the extensive description in the references [70] and [71].

## Results

We present the obtained *n*-dependence of the experimentally determined electro-kinetic mobilities for 19 different *n* in Fig. 4. We explicitly note that all samples with $n \geq 2\times10^{16}$ m$^{-3}$ displayed substantial multiple scattering, which we, however, could correct for using the improved version of SH-LDV. Displayed data are averages over several independent conditioning runs for each particle concentration. The electro-phoretic mobility increases from $\mu_{ep} = 3.0\times10^{-8}$ m$^2$V$^{-1}$s$^{-1}$at n = $3\times10^{15}$ m$^{-3}$ to $\mu_{ep} = 5.3\times10^{-8}$ m$^2$V$^{-1}$s$^{-1}$ at $n = 2\times10^{16}$ m$^{-3}$, then it plateaus up to n = $8.3\times10^{16}$ m$^{-3}$ from where it decreases again. In fluid samples, we also obtained the electro-osmotic mobility. Practically independent of the particle concentration, it stays constant at values close to $\mu_{eo} \approx 7.2\times10^{-8}$ m$^2$V$^{-1}$s$^{-1}$. As expected, neither mobility is affected by the volume phase transition from fluid to bcc, similar to the behavior of the conductivity. One observes a small, but non-systematic scatter in the osmotic mobilities, which we attribute to the adsorption of colloids on the cell walls. We support this point by additional measurements with different freshly burned and cleaned quartz cells. In this case, the mobility reaches a limiting value of $\mu_{eo} = 8.5\times10^{-8}$ m$^2$V$^{-1}$s$^{-1}$ (dashed line in Fig. 4) independent of the cell thickness.

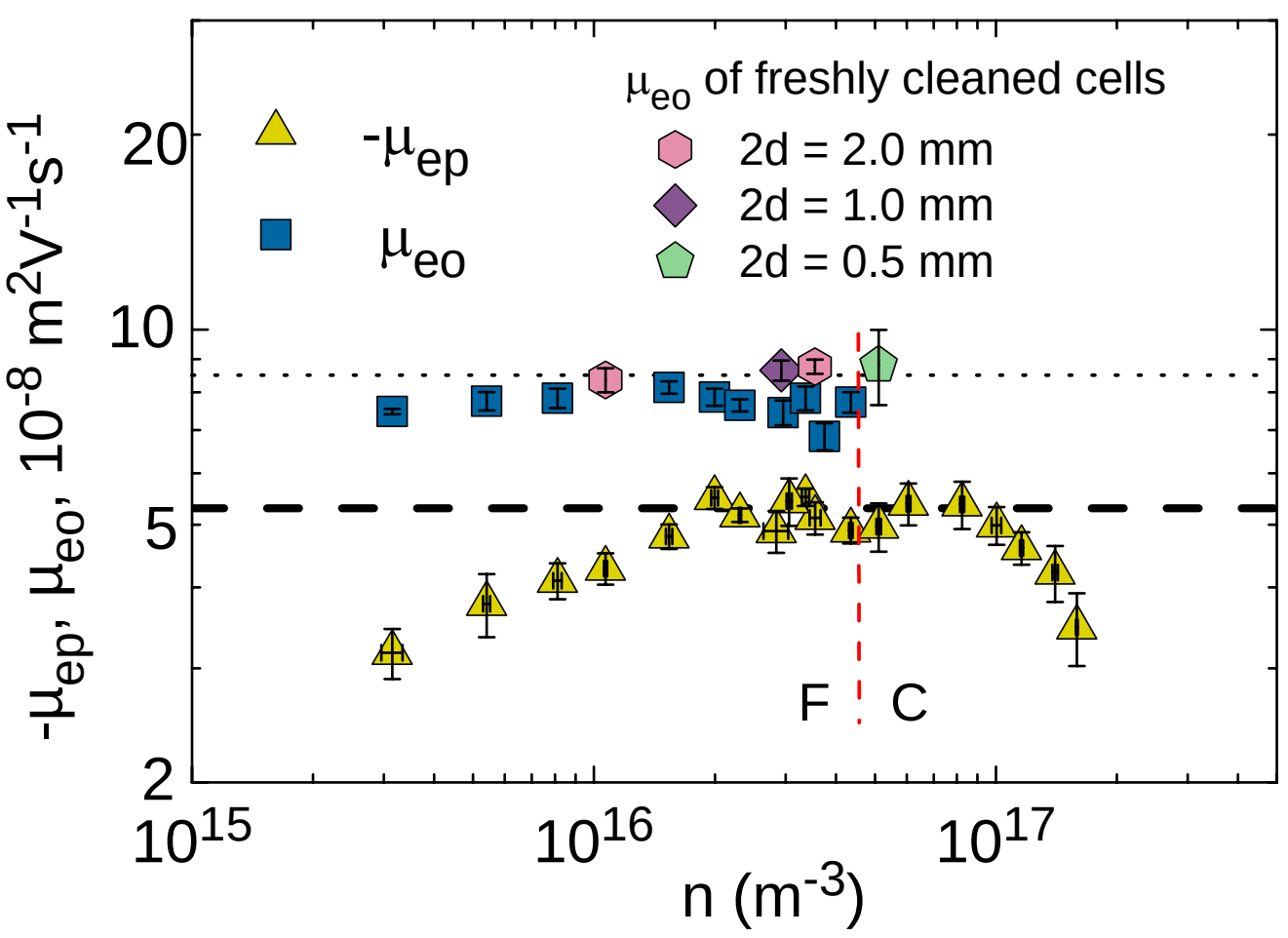

Fig. 4: Electro-kinetic mobilities as a function of *n*. The vertical dashed line indicates the fluid/crystal phase transition (F/C). The plateau value of electro-phoretic mobilities (yellow triangles) is $\mu_{ep} = 5.3\times10^{-8}$ m$^2$V$^{-1}$s$^{-1}$ (thick dashed line). The electro-osmotic mobility at the quartz cell (blue squares) wall is independent of particle concentration at a value of $\mu_{eo} \approx 7.2\times10^{-8}$ m$^2$V$^{-1}$s$^{-1}$. Freshly burnt cells show slightly larger values independent of cell thickness (coloured symbols). The limiting electro-osmotic mobility is of $\mu_{eo} = 8.5\times10^{-8}$ m$^2$V$^{-1}$s$^{-1}$ (thin dotted line). All samples with $n \geq 2\times10^{16}$ m$^{-3}$ showed substantial multiple scattering

We next compare the experimental electro-phoretic mobilities to the results of our calculations using the Standard Electro-Kinetic Model under realistic salt free conditions. For these, ambient carbon dioxide is allowed to dissolve as carbonic acid, which in turn is subject to a dissociation equilibrium. In addition, we consider both water hydrolysis and particle counter-ions to contribute to the overall electrolyte content. We show results for two limiting cases of $c(CO_2$, g) = $6.5\times10^{-8}$ mol L$^{-1}$ (corresponding to the minimum conductivity obtained experimentally in pure water) and $c(CO_2$, g) = $1.18\times10^{-5}$ mol L$^{-1}$ (corresponding to the saturation value). We calculated the density dependent electro-phoretic mobility for latex particles of $2a_h$ = (268±2) nm for these limiting cases. The colloid effective charge $Z_{eff} = -2500$ was taken from the conductivity measurements at elevated *n*. Fig. 5 compares both predictions to the experimental data on PS260(mP).

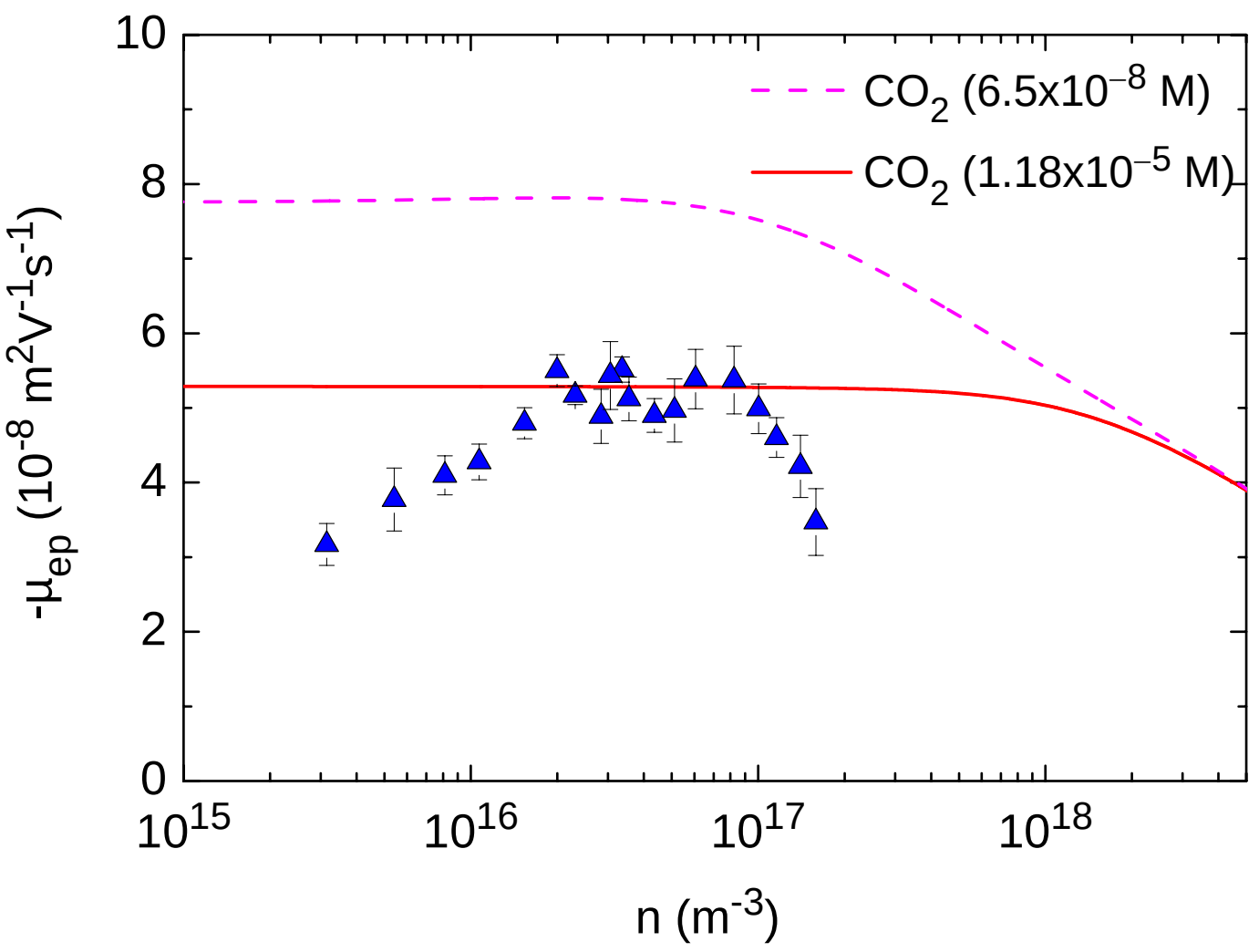

Fig. 5: Analysis of electro-phoretic mobilities for PS260(mP). Experimental $\mu_{ep}$ are compared to two Standard Electro-Kinetic Model calculations using the effective charge from conductivity ($Z_{eff} = -2500$) and the particle radius $2a_h$ = (268±2) nm as input. The concentrations of dissolved $CO_2$ are set to the value derived from minimal conductivity of water in our circuit (dashed) and to the saturation value at 25.2°C (solid). The plateau region of the experimental mobility is met by the second prediction. Experimental mobilities drop below the prediction on both sides of the plateau.

The second calculation coincides favorably with the values of the plateau region, while the low salt calculation predicts a much larger mobility. This on one side corroborates the finding that the particles themselves act as $CO_2$-reservoir. On the other side it demonstrates the consistency of conductivity and electro-phoretic experiments at elevated particle concentrations and $CO_2$-saturation across the plateau region. However, the experimentally observed decrease towards large *n* does not meet the constant charge predictions. For the present particles, the observed decrease starts at an order of magnitude lower concentration than expected particle concentrations. Moreover, it appears to be considerably faster. Also on the low *n* side, the experimental mobility values decrease, while the constant charge calculations expect a constant mobility.

## Discussion

We first compare our experimental findings to selected previous results obtained under similarly well-controlled realistic salt free conditions. In fact, the electro-phoretic behavior of charged spheres, shown in Fig. 4, was partially described before in several studies. Indications of a maximum were obtained early [17]. Indirect evidence of an increase with *n* was first reported by Deggelmann et al. [73]. The lack of detail as to both the latex cleaning and the interpretation of the mobility makes it difficult to discuss these data in relation to the present work. The low-*n*-increase of $\mu_{\text{ep}}$ with *n* was unequivocally demonstrated by Bellini et al. [30] for a low salt perfluorinated latex system and later reported also for polystyrene spheres [24] under carefully controlled salt concentrations. In both studies the investigated range of *n* was restricted due to sample crystallization. An increase was also seen at elevated salt concentrations in suspensions of small silica particles [31] and of surfactant stabilized latex spheres [28]. Medebach et al. [19] gave the first systematic report of a decrease of mobility with *n* for spheres with smaller radii under well controlled salt concentrations. Here insufficient single to noise ratio set a lower bound to the investigated range of particle concentrations. The decrease was later confirmed for spheres of about double that size [20]. A clear plateau-type maximum was first observed by Botin *et al.* [22]. However, also this latter study was performed only for a limited number of different *n* and without conductometric control. Fig. 6 shows a compilation of selected previous results for systems with precisely determined electrolyte content [19, 20, 24] together with the present data for PS260(mP). We note, that data taken under less quantified conditioning are excluded, but show qualitatively equivalent behavior [22, 29, 31, 28, 73]. The data compilation in Fig. 6 covers more than six orders of magnitude in *n*, some four orders of magnitude in Φ, more than one order of magnitude in κ*a*, and about six orders of magnitude in the counter-ion fraction $f = c_{H+} / c_{total}$. Following IUPAC recommendations, we display mobilities in reduced units to eliminate the influence of solvent viscosity $\eta_s$, relative permittivity, $\varepsilon_{rs}$ and temperature *T*: $\mu_{red} = (3/2)\,\mu\,\eta_s\,e/\varepsilon_0\,\varepsilon_{rs}\,k_B\,T$ [1].

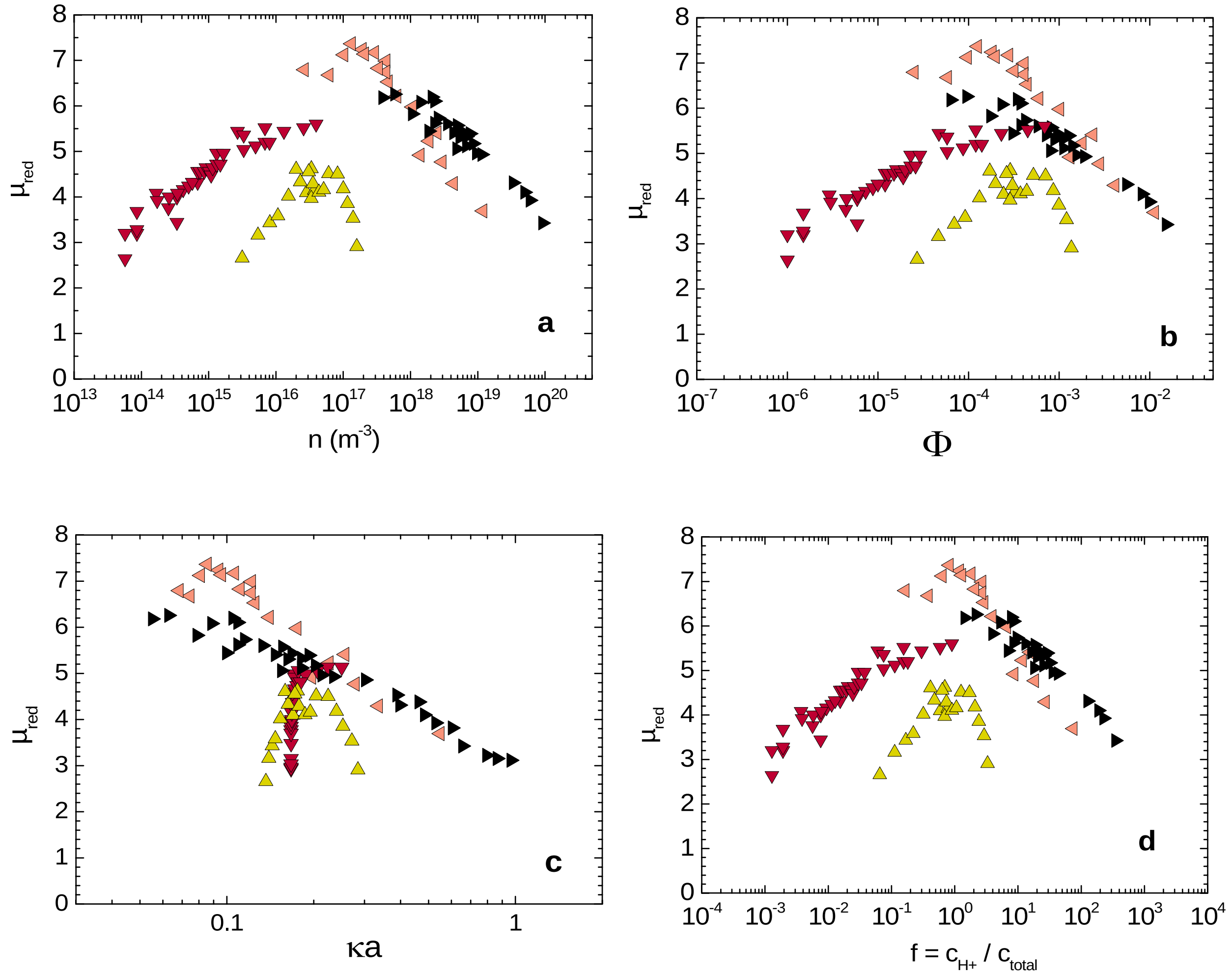

Fig. 6 Dependence of the reduced electro-phoretic mobility on different parameters for four selected particle systems. a) Semi-log plot of $\mu_{red}$ against number density. b) Semi-log plot of $\mu_{red}$ against volume fraction $\Phi = n(4\pi/3)a^3$. c) Semi-log plot of $\mu_{red}$ against reduced particle size $\kappa a$. d) Semi-log plot of $\mu_{red}$ against counter ion fraction $f = c_{H+} / c_{total}$. Symbols denote the following systems. Salmon left triangles: poly(n-butyl acrylate)-polystyrene (PnBAPS122, BASF, Ludwigshafen, Germany), $2a_h$ = (122±2) nm, $Z_{eff}$ = –750 [20]; black right triangles: poly(n-butyl acrylate)-polystyrene (PnBAPS68, BASF, Ludwigshafen, Germany), $2a_h$ = (68±3) nm, $Z_{eff}$ = –450 [19]; red down triangles: sulfate Polystyrene (PS, IDC, Portland Oregon) PS301, $2a_h$ = (322±4) nm $Z_{eff}$ = –2440 [24]; yellow up triangles: carboxylate modified Polystyrene (PS260(mP), microParticles GmbH, Berlin, Germany) $2a_h$ = (269±6) nm, $Z_{eff}$ = –2500, this study.

In Fig. 6, only the data set of PS260(mP) displays a defined plateau type maximum as a function of concentration. We can demonstrate this remarkable behavior here for the first time for a single well-characterized particle species under carefully controlled conditioning. The other curves show either a plateau followed by a descent or an ascent followed by a plateau. Our new data, therefore, qualitatively reconcile the conflicting previous results on possible types of particle concentration dependence. In fact, they show three distinct regions of increase, plateau and decrease, which in the other data sets obtained on restricted concentration ranges are present only partially. Our observation further corroborate other earlier results obtained under less precise control of boundary conditions [22, 29, 29, 31, 28, 73] as well as studies reporting very low mobilities for very low volume fractions or isolated particles [32, 33, 34, 35].

Closer inspection, however, reveals some significant differences between different particle species. Both the number density (Fig. 6a) and the volume fraction (Fig. 6b) of the transition between the different behaviors differ considerably, sometimes by more than an order of magnitude. Also the ranges, over which theoretical constant charge calculations or primitive level simulations describe the data well differ considerably. For PnBAPS68 and PnBAPS122 an overall good agreement was observed throughout plateau and descent [15, 20, 21]. For PS301 [24] and for PS260(mP) this only holds in the plateau region, i.e. over roughly one order of magnitude in concentration.

These differences also are obvious from plotting the results versus other parameters. A traditional way of plotting electro-phoretic data is versus the reduced particle radius $\kappa a$, where $\kappa$ is the Debye-Hückel screening parameter $\kappa = \sqrt{e^2 / \varepsilon_0 \; \varepsilon_{rs} \; k_B T \sum_k n_k \; z_k}$. In Fig. 6c we calculated $\kappa$ accounting for all micro-ionic species $k$ present, i.e. we include releases counter-ions, ions from solvent hydrolysis and from $CO_2$ dissolution. The two smaller PnBAPS species show a stretched descent reflecting the gradual increase of screening upon adding more and more counter-ions. This reflects the known salt concentration dependence of the mobility at constant particle charge or $\zeta$ potential [1, 8, 15]. By contrast, PS301 and PS260(mP) change their mobility at hardly changing the electrolyte content i.e. the screening conditions. Therefore, their abrupt mobility change reflects a change of their $\zeta$ potential and their particle charge, respextively. Interestingly, for both species, the transition from ascent to plateau is found to coincide fairly well with the completion of double layer overlap, i,e, $\kappa n^{-1/3} \approx$ 3-7. In Fig. 6d we follow [31] and plot the data versus the electrolyte composition in terms of the counter-ion fraction $f = c_{H+} / c_{total}$. This quantity reflects two distinct effects. First at $f > 1$, counter-ions are dominating the screening behavior and the salinity gradually increases with increasing particle concentration. This is seen for PnBAPS68 and PnBAPS122 with the onset located at $f$-values around 2-5. Second, for $f > 1$, the counter-ions also start influencing the suspension pH while for $f < 1$it either stays constant or is controlled by the buffer function of the $CO_2$ dissociation equilibrium. The steep descent in the data for PS260(mP) starting at $f \approx 2$ appears to be compatible with a pH triggered decrease. By contrast, the ascent of PS301 and PS260(mP) both commence already at very low values of $f$ ruling out a connection to screening and pH. Our comparison shows that while we have observed this behavior in experiments conducted in dependence on concentration, the influence of particle concentration appears to be indirect. Nor does the freezing transition interfere. For PS260(mP) $n_F = (4.6 \pm 0.2) \times 10^{16}$ m$^{-3}$ is in the center of the plateau region, for PnBAPS68 and PnBAPS122, $n_F = 4 \times 10^{17}$ m$^{-3}$ and $n_F = 6 \times 10^{18}$ m$^{-3}$ are halfway down the descent.

Keeping this in mind, we return to Fig. 5 and take a closer look at the comparison of the present experimental and theoretical results. Mobilities measured on PS260(mP) agree quantitatively across the plateau region with the predictions based on the experimental effective charge from conductivity. A constant mobility is expected from our theoretical calculations under realistic salt free conditions as long, as the effective charge stays constant and the salinity is dominated by the added electrolyte. From the onset of counter-ion dominance, the mobility decreases with increasing *n*, as can be seen on the right side of the theoretical curves in Fig. 5. The mobility of PS260(mP), however starts to drop already at lower concentration and in a much steeper way. In the calculations, we used the conductivity effective charge for all particle concentrations. This is justified, as within charge renormalization theory, no significant net effect on $Z_{eff}$ is expected from increasing *n* [12]. Over the complete experimental range, the slight increase in $Z_{eff}$ with decreasing $\kappa$ is counteracted by an equally mild increase in *A* [36]. On the other side, with calculations based on the renormalized charge, there is hardly any way within the Standard Electro-Kinetic Model to reduce $\mu_{ep}$ in the abrupt ways seen in the experiment. Any significant drop of mobility there corresponds to a strong decrease of the ζ-potential and the effective charge, respectively. This in turn points to a significant reduction of bare charge such as to leave the saturation limit of charge renormalization [10, 12]. From this we conclude that in the present case: *i*) charge renormalization applies; *ii*) ζ-potential and effective charge are constant across the plateau; *iii*) ζ-potential and effective charge drop quickly both towards smaller and towards larger *n* which points at decrease of the bare charge. The presence of these pronounced deviations shows that the commonly made constant charge assumption in interpreting charged sphere suspension data should be taken with due caution.

At large *n*, a decrease of bare charge is readily anticipated considering, that the only counter-ions in our system are $H^+$. Their accumulation in the electric double layer changes the surface pH and thus can influence the dissociation equilibrium for surface groups of sufficiently large $pK_A$. This effect, known as charge regulation, in fact is a straightforward implementation of chemical equilibrium in the calculation of the double layer properties [74, 75]. Experimentally it has been demonstrated from colloidal probe AFM [76, 77, 78], but also from electro-kinetics of non-polar colloids [79]. From Fig. 6d, we noted that the onset of the mobility drop occurs at $f \approx 2$, i.e. just when the $H^+$ counter-ions start influencing the suspension pH. We therefore propose that the experimentally observed drop in mobility at large *n* corresponds to a pronounced decrease of bare charge caused by charge regulation, *i.e*. a decrease in the degree of surface group dissociation. This effect could be implemented into the present Standard Electro-Kinetic Model approach by the numerically efficient *ab initio* calculations of colloidal bare and effective charges under realistic salt free conditions that were recently introduced for sulfonate and silanol stabilized particles [13]. In turn, charge regulation based Standard Electro-Kinetic Model calculations could be performed and compared to the experimental data.

Also at low *n*, the qualitative difference in the observed electro-kinetic behavior for the two different surfaces investigated in this study supports the idea of a physicochemical equilibrium type mechanism. In fact, the electro-osmotic mobility at the Quartz surface stays constant, and the corresponding averaged $\zeta$-potential is $\zeta_{Quartz} = -125$ mV independent of *n* and in good agreement with previous measurements [80]. The $\zeta$-potential of freshly burnt cells amounts to $\zeta_{Quartz} = -141$. At the same time the mobility at the PS-surface drops by about a factor of two. However, simple pH-controlled charge regulation as sketched above can presumably be ruled out, since it is unclear, where the required excess in $H^+$ should come from for decreasing *n*. Thus, even while a decreasing bare charge is suggested from the comparison to the Standard Electro-Kinetic Model, the underlying mechanism remains unresolved for the low *n* side.

At this point we feel free to speculate and anticipate two lines of thought that might be worth considering. The first line of thought is based on the salt concentration dependence of the solubility of neutral $CO_2$ in water. This dependency has been thoroughly studied for sea-water due to its relevance for climate issues [81, 82]. In a recent study, we further investigated the $CO_2$ solubility in low salt water and, in line with the well-established findings at large overall salinity, found it to further decrease upon deionization [83]. In a thought experiment, we start from a suspension of arbitrary particle concentration with the solvent at equilibrium with ambient air. We then deionize the suspension. This decreases the electrolyte content (residual salt and dissociated carbonate ions). It does, however not decrease the content of $CO_2$ dissolved as neutral molecule. As long as the number density stays large enough, the electric double layers are overlapping. For $\kappa n^{-1/3}/2 > 1$ regions of “pure” solvent appear. The storage capacity of the double layer regions remains unaltered. In the ”pure” solvent, a reduced salinity reduces the saturation concentration of neutral $CO_2$ [81, 82], such that upon deionization at low enough *n*, the suspension becomes over-saturated in dissolved neutral $CO_2$. Dissociation after cessation of deionization only partially solves the problem. The surplus $CO_2$ then either leaves by bubble formation (as reported in [83]) or adsorb to soft low dielectric particle surfaces to form nano-bubbles [40]. At constant number density, less $CO_2$ per particle will be stored on larger particles. With decreasing particle number density, the amount stored per particle should increase. As a consequence of the increased surface concentration of $CO_2$, the local dielectric constant is decreased, and the enhanced electrostatic interaction between surface charges can lead to a shift of their chemical dissociation equilibrium [74, 75]. Therefore, we tentatively suggest as a possible mechanism for the low-*n* behaviour of the electro-phoretic mobility charge regulation triggered by a change in dielectric permittivity.

Alternatively, one may suspect a lowered dielectric constant to favor the build-up of correlations amongst counter-ions and also between surface charges and counter-ions [84]. A first type of correlations has been investigated intensively for systems with smeared surface charge and finite sized counter-ions [85, 86]. It becomes most relevant at large concentrations of electrolyte. A second type of electro-static correlation leads to Bjerrum-pair formation and can, in the case of di- or multivalent counter-ions, even invert the particle charge [87]. For the present case of monovalent counter-ions, we expect this physical effect to decrease the particle charge with decreasing dielectric constant, *i.e.* increasing $CO_2$ adsorption.

Such a mechanism appears to be qualitatively consistent with the data in Fig. 6a. For both PS301 and PS260(mP) we note that the plateau region ends towards smaller *n* upon the loss of double layer overlap at $\kappa n^{-1/3}/2 \approx 1.5$ and $\kappa n^{-1/3}/2 \approx 3.5$, respectively. Further, PS301 shows a smaller slope than PS260(mP). This could, however also relate to a different surface $pK_A$. This suggestion could be further tested by careful systematic measurements of $CO_2$ solubility under low salt conditions and further electro-kinetic experiment with particles of differing materials to vary the adsorption equilibrium.

A different line of thought is related to the similarity between deionization and degassing. As our conductivity measurements show, the amount of dissolved $CO_2$ is greatly reduced upon deionization over extended times. Under degassed conditions a change of the surface structure has been observed which increases the surface roughness [42]. From investigations of “hairy surface” particles it is further known, that surface roughness may considerably reduce the electro-kinetic mobility [88]. Force measurements in degassed water and surface roughness characterization by AFM comparing as-synthesized to deionized particles may address these suggestions. This mechanism can be experimentally discriminated from those mechanisms based on $CO_2$ trapping, as it would affect mainly viscoelastic surfaces and leave Silica surfaces unaltered. This is in line with our observations on the quartz cell walls, but remains to be tested for particles with rigid surfaces.

## Conclusion

We have presented a first systematic investigation of charged sphere electro-phoretic mobilities covering a sufficiently large range of concentrations to unequivocally identify a plateau-type mobility maximum. Data were taken on highly dilute, fluid and crystalline ordered samples. Most importantly, we could include samples, where multiple scattering effects occur and render the systems too turbid for standard instrumentation even at volume fractions of about 1%. We applied a facile multiple scattering correction to our SH-LDV data to obtain high quality Doppler spectra for very turbid systems of low transmittance. Our data qualitatively reconcile conflicting previous results on a density dependence of the electro-phoretic mobility taken over restricted density ranges. They thus solve a long standing *experimental* challenge.

Our comparison to the predictions of the Standard Electro-Kinetic Model emphasized the importance of considering realistic salt free conditions. Thus, our predictions for $CO_2$ saturated conditions meet the experimental data of the plateau region quantitatively and without additional adjustable parameters when taking the independently determined effective charge as input. On one side, this finding is in line with the conductivity measurements, which reveal that the particles themselves act as a non-negligible source of $CO_2$. On the other side, it may also be seen as a stringent test of the theoretical approach. The Standard Electro-Kinetic Model works very well under realistic salt free conditions, given the experimental systems have constant charges.

From the data obtained over a sufficiently large concentration range, as well as from their comparison to the theoretical calculations and to selected previous results obtained under similarly well-defined conditions, we could further discriminate three distinct regions of differing density dependence for our particles. In particular, we found an ascent during the built up of double layer overlap which levels off to a plateau. Such an ascent is not anticipated by the Standard Electro-Kinetic Model based on a constant effective charge. The mobilities across the extended plateau region are, however, in quantitative agreement with the expectation. The steep and sudden decrease differs significantly from the gradual mobility decrease due to counter-ion dominated screening which is found by theory and simulation, and which was previously seen for other particle species.

The peculiar behavior of the mobility towards very small and very large densities seen in the present study is not expected from constant charge calculations. Rather, the observed drop points at a reduced bare charge. Closer analysis showed that different mechanisms might apply in different concentration ranges. Each of the tentatively suggested mechanisms relies on an indirect influence of the number density and its complex interplay with other experimental parameters and the particle surface properties. It will therefore be rewarding to extend the present Standard Electro-Kinetic Model calculations, e.g. by implementing surface charge regulation, and compare them to future experiments at elevated *n*. Additional experiments are also desired at low *n*, to explore the range of concentrations with and without double layer overlap in more detail. Apart from number density, variation of particle surface chemistry (e.g. particles bearing weak acids of different $pK_A$), surface rigidity and, perhaps most importantly, a controlled variation of the $CO_2$ concentration should be addressed. Solving the puzzle of a density dependence of the particle charge is highly desired for many practical applications of colloidal suspensions requiring systems at different particle concentrations. We anticipate that multiple scattering corrected SH-LDV will provide much support for such studies over extended concentration ranges by providing a solid database for further theoretical modelling.

## Acknowledgements

We like to thank A. Delgado, R. Roa, Ch. Holm, V. Lobashkin, M. Heinen, G. Nägele, B. Marino, G. Trefalt and M. Borkovec for their continued interest in and patience with our work resulting in many intense discussions. Financial Support of the DFG (Grant no. Pa459/18-1,2) is gratefully acknowledged. Partial financial supports for this work by the Consejería de Conocimiento, Investigación y Universidad, Junta de Andalucía and European Regional Development Fund (ERDF), ref. SOMM17/6105/UGR, as well as computer resources provided by Institute Carlos I for Theoretical and Computational Physics (IC1), University of Granada, are gratefully acknowledged.

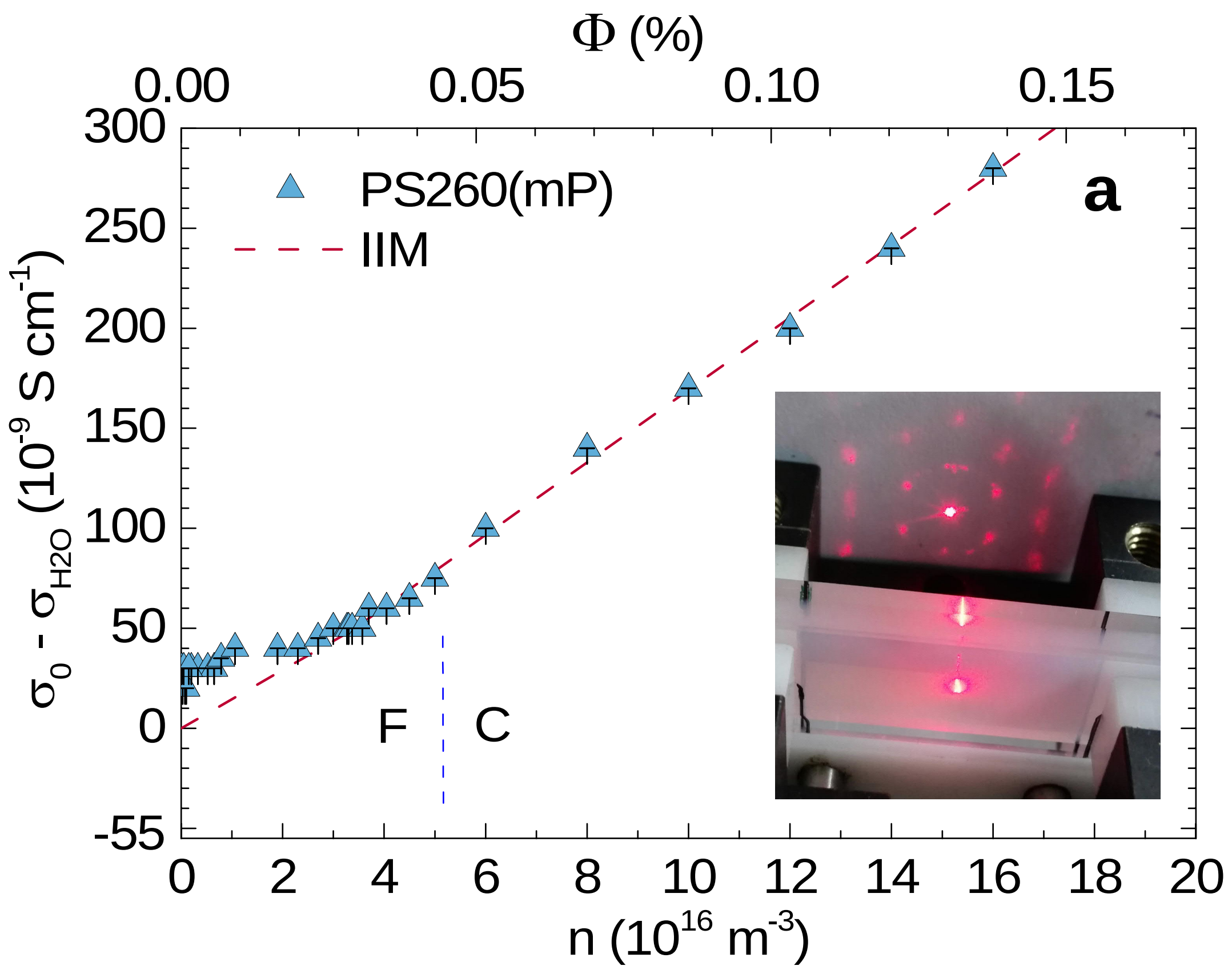
Φ (%)
0.00
0.05
0.10
0.15
PS260(mP)
IIM
a
σ0 - σH2O (10-9 S cm-1)
F
C
n (1016 m-3)

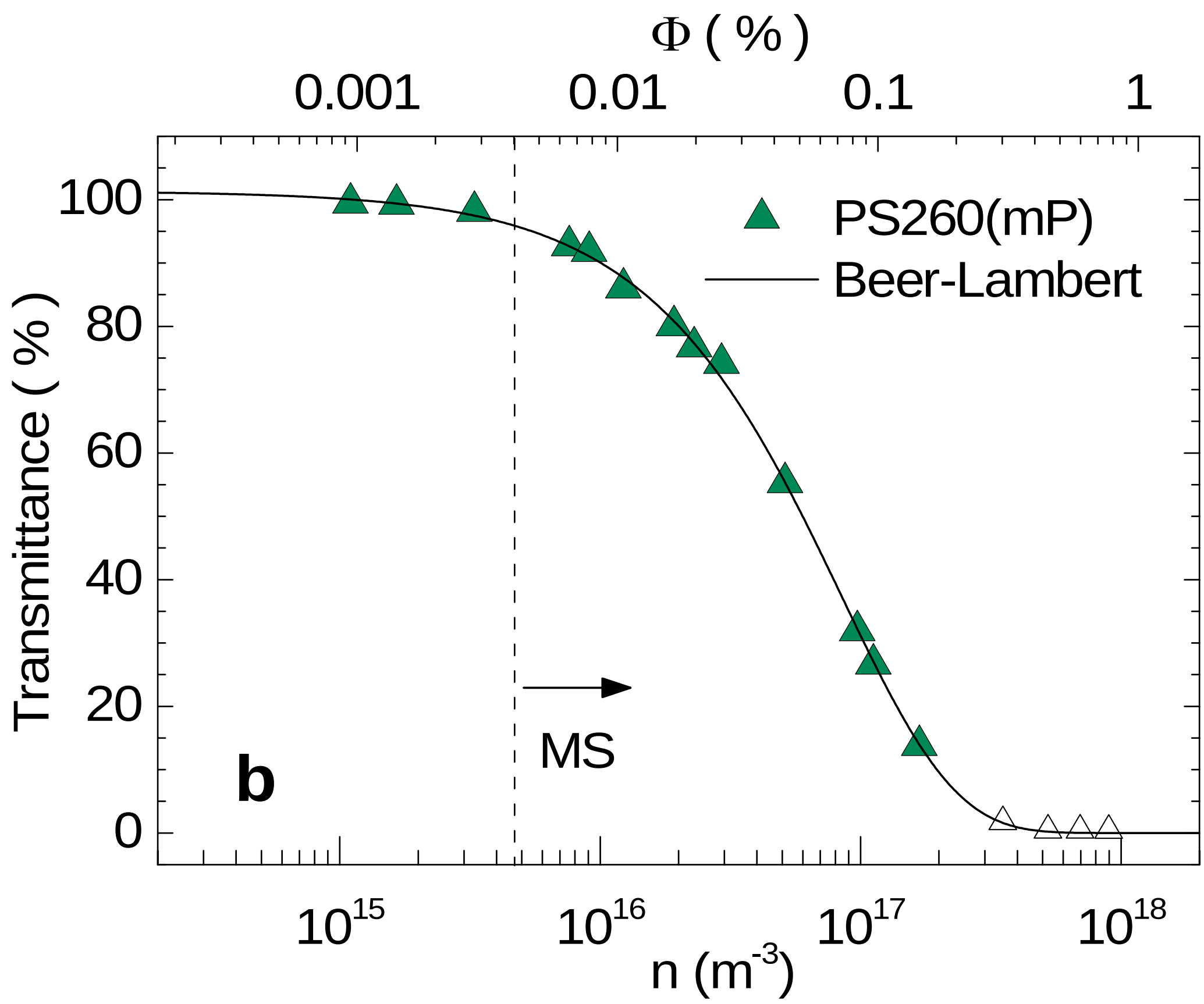
Φ ( % )
0.001
0.01
0.1
1
PS260(mP)
Beer-Lambert
Transmittance ( % )
MS
b
n (m-3)

**Fig. 1 ab**

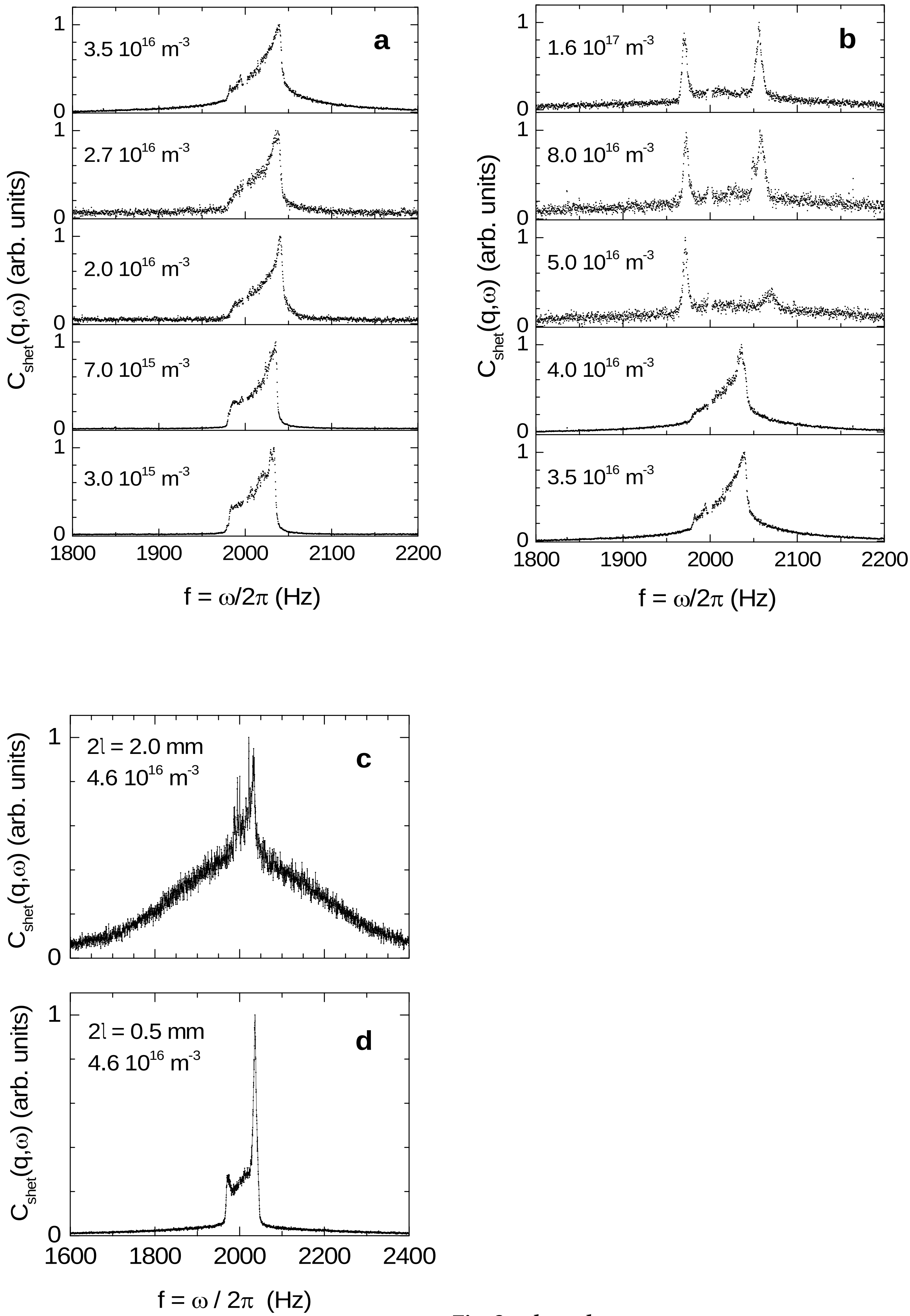
a
3.5 10^16 m^-3
2.7 10^16 m^-3
2.0 10^16 m^-3
7.0 10^15 m^-3
3.0 10^15 m^-3
b
1.6 10^17 m^-3
8.0 10^16 m^-3
5.0 10^16 m^-3
4.0 10^16 m^-3
3.5 10^16 m^-3
C_shet(q,ω) (arb. units)
f = ω/2π (Hz)
1800
1900
2000
2100
2200
c
2l = 2.0 mm
4.6 10^16 m^-3
d
2l = 0.5 mm
4.6 10^16 m^-3
1600
1800
2000
2200
2400
f = ω / 2π (Hz)
0
1

**Fig. 2 a, b, c, d**

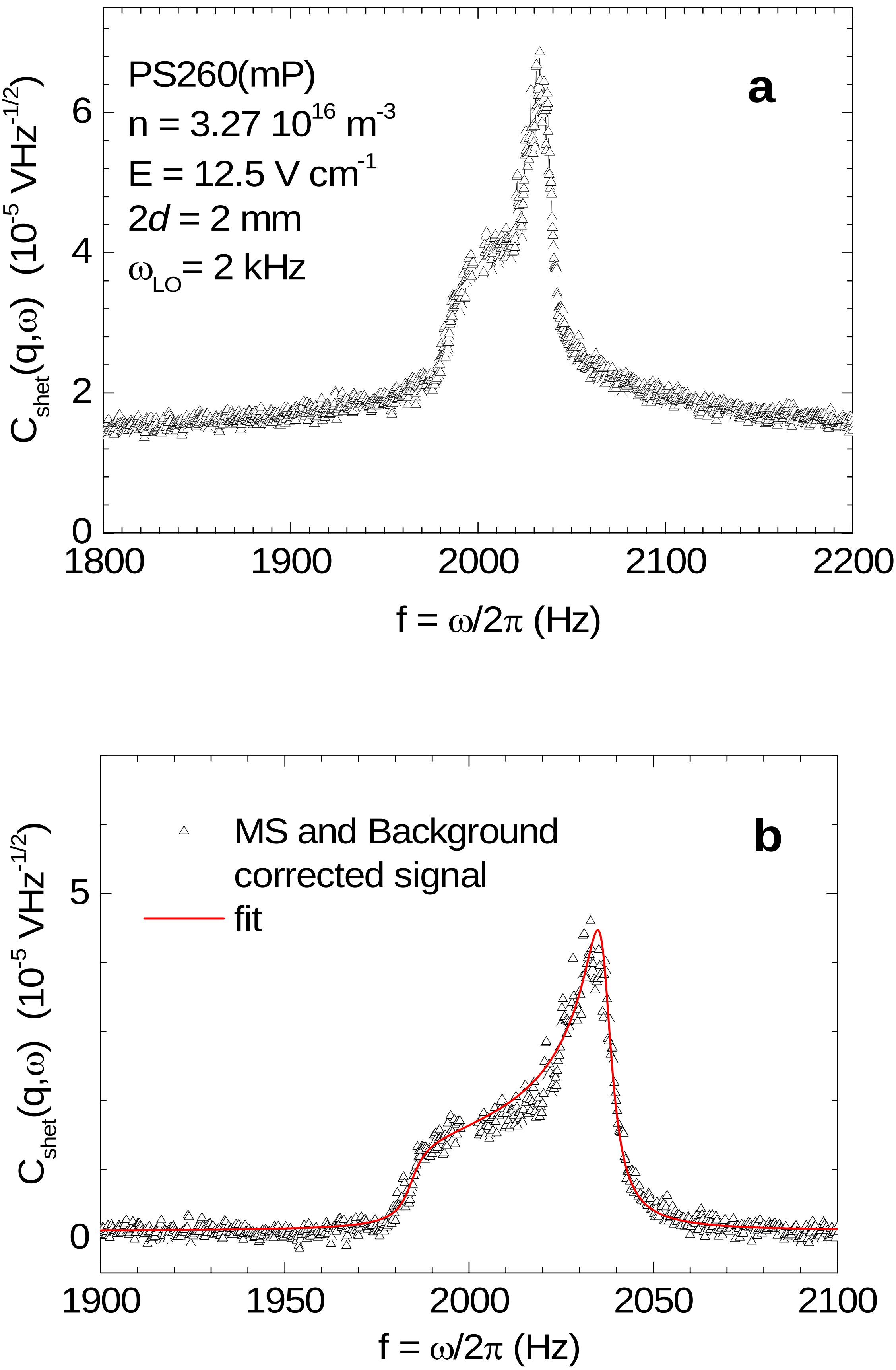
PS260(mP)
n = 3.27 10^16 m^-3
E = 12.5 V cm^-1
2d = 2 mm
ω_LO = 2 kHz
a
C_shet(q,ω) (10^-5 VHz^-1/2)
f = ω/2π (Hz)
MS and Background
corrected signal
fit
b
C_shet(q,ω) (10^-5 VHz^-1/2)
f = ω/2π (Hz)

**Fig. 3a, b**

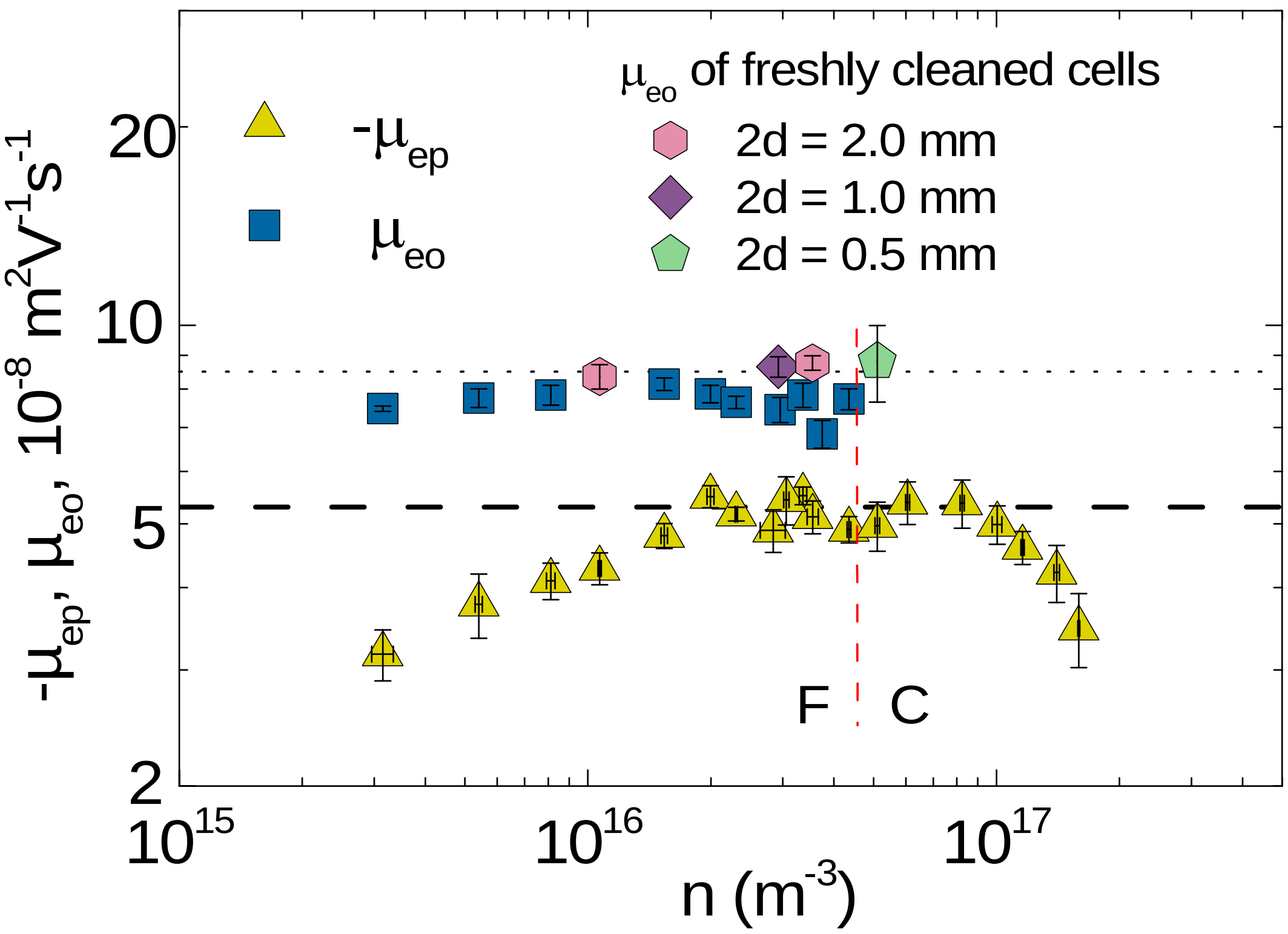

-μep
μeo
μeo of freshly cleaned cells
2d = 2.0 mm
2d = 1.0 mm
2d = 0.5 mm
F
C
-μep, μeo, 10-8 m2V-1s-1
n (m-3)

**Fig. 4**

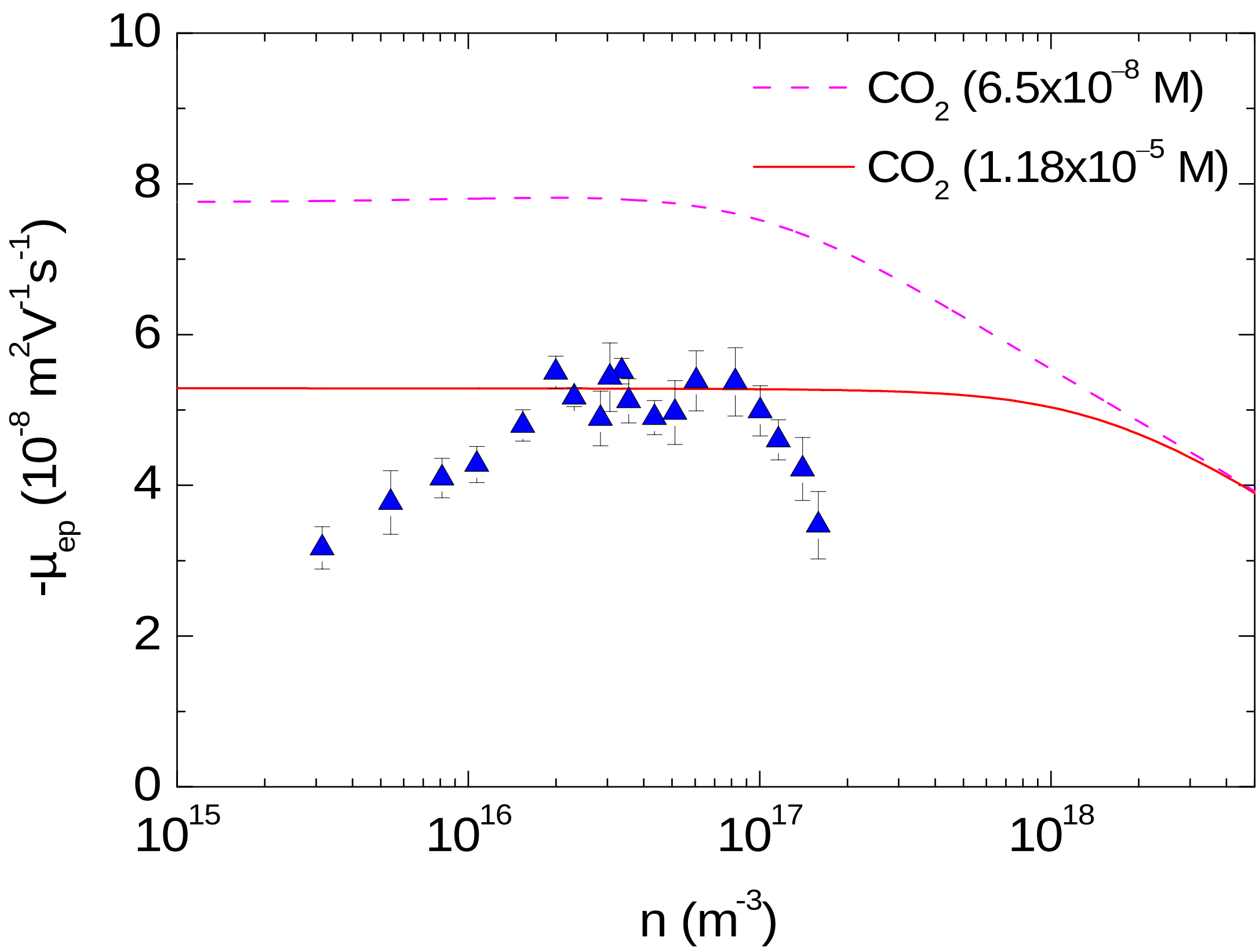

CO2 (6.5x10-8 M)
CO2 (1.18x10-5 M)
-μep (10-8 m2V-1s-1)
n (m-3)

**Fig. 5**

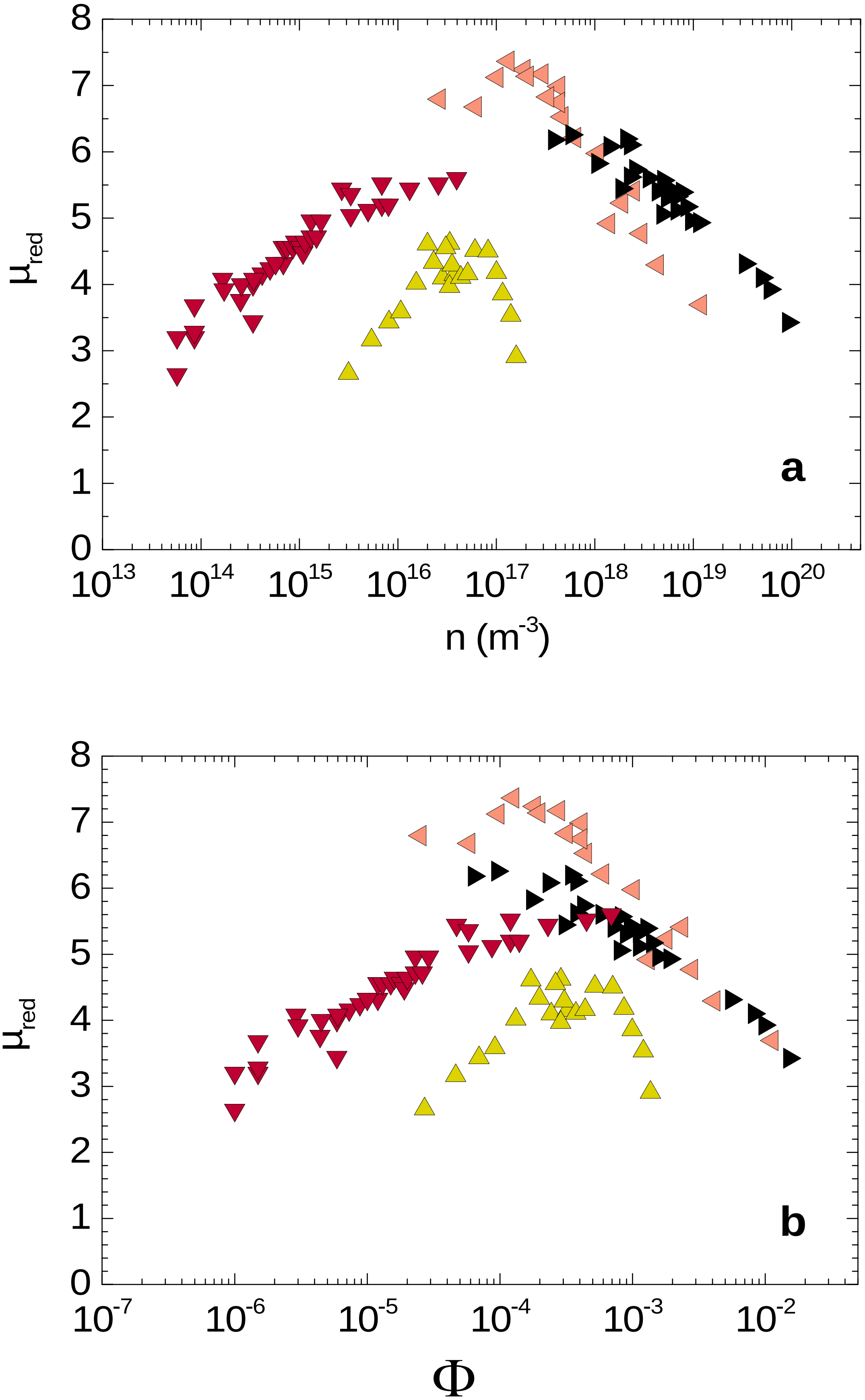

8
7
6
5
4
3
2
1
0
μred
10¹³
10¹⁴
10¹⁵
10¹⁶
10¹⁷
10¹⁸
10¹⁹
10²⁰
n (m⁻³)
a
8
7
6
5
4
3
2
1
0
μred
10⁻⁷
10⁻⁶
10⁻⁵
10⁻⁴
10⁻³
10⁻²
Φ
b

**Fig. 6a, b**

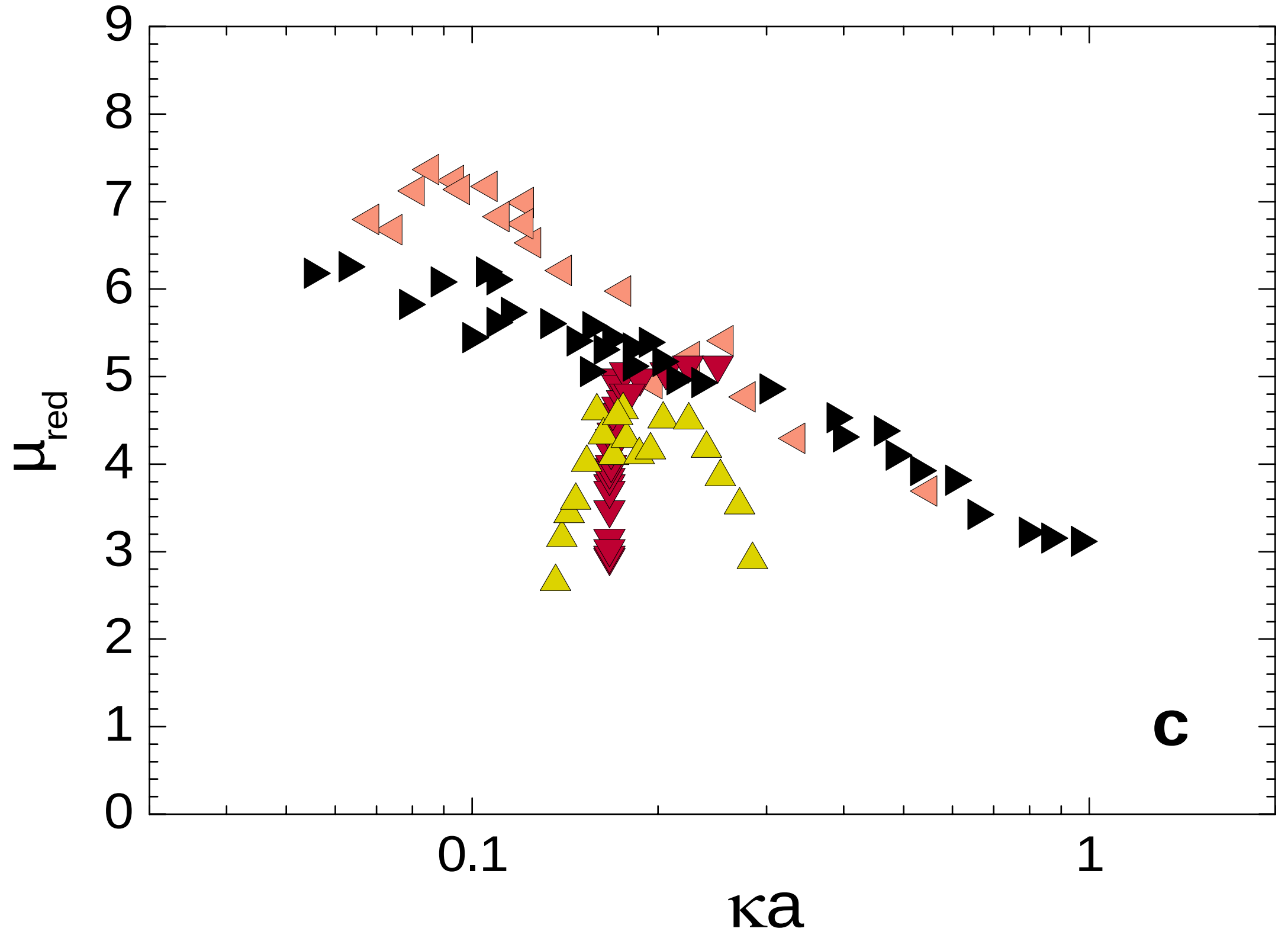
9
8
7
6
5
4
3
2
1
0
μred
c
0.1
1
κa

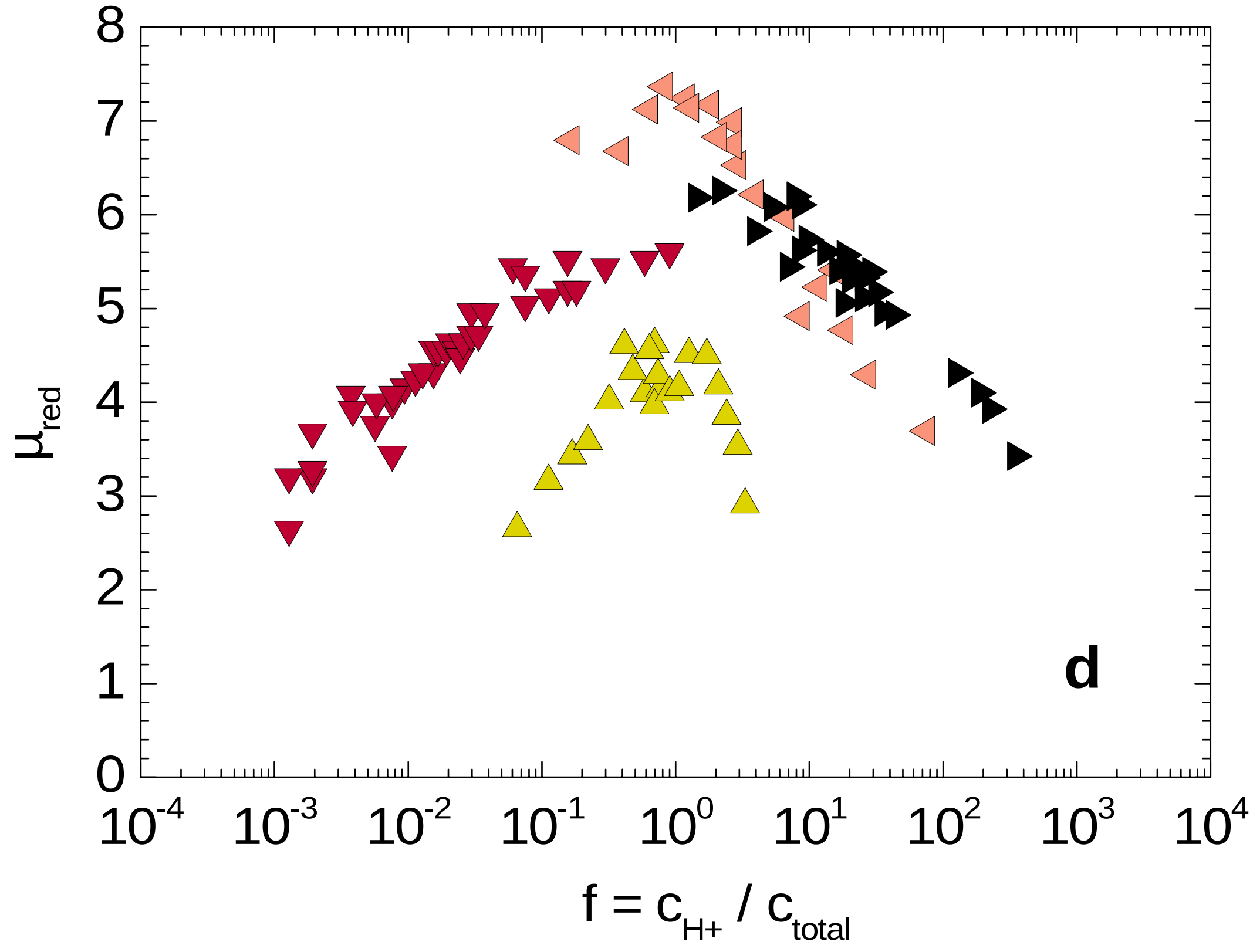
8
7
6
5
4
3
2
1
0
μred
d
10-4 10-3 10-2 10-1 100 101 102 103 104
f = cH+ / ctotal

**Fig. 6c, d**